\documentclass[a4paper,12pt]{article}
\usepackage{amsmath}
\usepackage{amssymb}
\usepackage{amsfonts}
\usepackage{amssymb,amsmath}
\usepackage{cancel}
\usepackage{graphicx}
\usepackage{epsfig}
\usepackage{verbatim}
\usepackage{fleqn}
\usepackage[footnotesize]{caption}
\usepackage{mathrsfs}
\usepackage{float}
\def\beq{\begin{equation}}
\def\eeq{\end{equation}}
\def\bea{\begin{eqnarray}}
\def\eea{\end{eqnarray}}

\def\<{\left\langle}
\def\>{\right\rangle}

\newcommand{\bc}{\begin{center}}
\newcommand{\ec}{\end{center}}
\newcommand{\bd}{\begin{displaymath}}
\newcommand{\ed}{\end{displaymath}}
\newcommand{\be}{\begin{equation}}
\newcommand{\ee}{\end{equation}}
\newcommand{\ba}{\begin{array}}
\newcommand{\ea}{\end{array}}
\newcommand{\bt}{\begin{tabular}}
\newcommand{\et}{\end{tabular}}

\newcommand{\ds}{\displaystyle}

\newcommand{\BC}{\begin{center}}
\newcommand{\EC}{\end{center}}
\newcommand{\BE}{\begin{equation}}
\newcommand{\EE}{\end{equation}}
\newcommand{\BEA}{\begin{eqnarray}}
\newcommand{\BEAnn}{\begin{eqnarray*}}
\newcommand{\EEA}{\end{eqnarray}}
\newcommand{\EEAnn}{\end{eqnarray*}}

\newcommand{\VL}{\left( \begin{array}{c}}
\newcommand{\VR}{\end{array} \right)}
\newcommand{\ML}{\left( \begin{array}{cc}}
\newcommand{\MLd}{\left( \begin{array}{ccc}}
\newcommand{\MLv}{\left( \begin{array}{cccc}}
\newcommand{\MR}{\end{array} \right)}

\newcommand{\sfl}{\tilde{f}_L}
\newcommand{\sfr}{\tilde{f}_R}
\newcommand{\sfe}{\tilde{f}_1}
\newcommand{\sfz}{\tilde{f}_2}

\newcommand{\tsf}{\theta\kern-.20em_{\tilde{f}}}
\newcommand{\tsfp}{\theta\kern-.20em_{\tilde{f}\prime}}
\newcommand{\tsq}{\theta\kern-.15em_{\tilde{q}}}

\newcommand{\costf}{\cos\tsf}

\newcommand{\sintf}{\sin\tsf}

\begin{document}

\bibliographystyle{OurBibTeX}

\begin{titlepage}

%\vspace*{-15mm}
\begin{flushright}
SHEP-10-42\\
\end{flushright}
%\vspace*{5mm}

\begin{center}
{ \sffamily \Large LHC Signatures of the \\[6mm]
Constrained Exceptional Supersymmetric Standard Model }
\\[8mm]
P.~Athron$^{a}$,
%\footnote{E-mail: \texttt{p.athron@physics.gla.ac.uk}}
S.F.~King$^{b}$,
%\footnote{E-mail: \texttt{sfk@hep.phys.soton.ac.uk}}
D.J.~Miller$^{c}$,
%\footnote{E-mail: \texttt{d.miller@physics.gla.ac.uk}}
S.~Moretti$^{b}$
%\footnote{E-mail: \texttt{stefano@phys.soton.ac.uk}}
and
R.~Nevzorov$^{d}$\footnote{On leave of absence from the Theory Department,
ITEP, Moscow, Russia.
%\qquad\qquad\qquad\qquad\qquad\qquad\qquad
%E-mail: \texttt{nevzorov@phys.soton.ac.uk}
}\\[3mm]
{\small\it
$^a$ Institut f\"ur Kern- und Teilchenphysik, TU Dresden, D-01062, Germany,\\[2mm]
$^b$ School of Physics and Astronomy, University of Southampton,\\
Southampton, SO17 1BJ, U.K.\\[2mm]
$^c$ SUPA, School of Physics and Astronomy, University of Glasgow,\\
Glasgow G12 8QQ, U.K.\\[2mm]
$^d$ 
Department of Physics and Astronomy, University of Hawaii, Honolulu, HI 96822
 }
\\[1mm]
\end{center}
\vspace*{0.75cm}

\begin{abstract}
\noindent

We discuss two striking Large Hadron Collider (LHC) signatures of the
constrained version of the exceptional supersymmetric standard model
(cE$_6$SSM), based on a universal high energy soft scalar mass $m_0$,
soft trilinear coupling $A_0$ and soft gaugino mass $M_{1/2}$.  The
first signature we discuss is that of light exotic colour triplet charge $1/3$
fermions, which we refer to as $D$-fermions. We calculate the LHC
production cross section of $D$-fermions, and discuss their decay
patterns. Secondly we discuss the $E_6$ type $U(1)_N$ spin-1 $Z'$
gauge boson and show how it may decay into exotic states, increasing
its width and modifying the line shape of the dilepton final state.
We illustrate these features using two representative cE$_6$SSM benchmark points, 
including an ``early LHC discovery'' point, giving the Feynman rules and numerical 
values for the relevant couplings in order to facilitate further studies.
\end{abstract}

\end{titlepage}
\newpage
\setcounter{footnote}{0}

\section{Introduction}

Last year the LHC experiments started to collect data. We expect that the LHC
will shed light on the physics beyond the Standard Model (SM), the
origin of dark matter and the mechanism of electroweak symmetry
breaking in the near future. However it may take some time for the
LHC experiments to discover the Higgs boson if it is light. On the other
hand the LHC can relatively quickly discover new coloured particles and a $Z'$
if these states are kinematically accessible. In this article we study the
production and decay signatures of the $Z'$ and exotic colour triplet charge $1/3$
fermions, which we refer to as $D$-fermions, that naturally
appear within well a motivated supersymmetric extension of the SM known as the 
Exceptional Supersymmetric Standard Model (E$_6$SSM) \cite{King:2005jy}.

Softly broken supersymmetry (SUSY) provides a very attractive framework 
for physics beyond the standard model (BSM), in which the hierarchy problem is 
solved and the unification of gauge couplings can be realised \cite{Chung:2003fi}. 
Despite these attractive features, the minimal supersymmetric standard model 
(MSSM) suffers from the $\mu$ problem.
%The minimal supersymmetric standard model (MSSM) \cite{Chung:2003fi}
%provides a very attractive scenario for physics beyond the standard model
%(SM), in which the hierarchy problem is solved and the unification of
%gauge couplings can be realised.  Despite these attractive features,
%the MSSM suffers from the $\mu$ problem.  
The superpotential of the MSSM contains the bilinear term $\mu
{H}_d{H}_u$, where ${H}_{d,u}$ are the two Higgs doublet
superfields\footnote{Note that we will not put hats on the
superfields.}  whose scalar components develop vacuum expectation
values (VEVs) at the weak scale and $\mu$ is the supersymmetric Higgs
mass parameter which can be present before SUSY is broken. One
naturally expects $\mu$ to be the order of the Planck mass or to be
zero, having been forbidden by some symmetry, whereas
phenomenologically, to achieve correct electroweak symmetry breaking
(EWSB), $\mu$ is required to be in the TeV region.

A very elegant solution to this problem is to generate an effective
$\mu$-term from an interaction, $\lambda S H_d H_u$, between the usual
Higgs doublets and a new Higgs singlet superfield $S$, whose scalar component develops a
low energy VEV.  However, although an extra singlet superfield
$S$  seems like a
minor modification to the MSSM, which does no harm to gauge
coupling unification, its introduction leads
to an additional accidental global $U(1)$ (Peccei-Quinn (PQ)
\cite{Peccei:1977hh}) symmetry which will result in a weak scale
massless axion when it is spontaneously broken by the VEV $\langle S\rangle$
\cite{Fayet:1974fj}.

To avoid this one can promote the PQ symmetry to an Abelian $U(1)'$
gauge symmetry \cite{Fayet:1977yc}.  The troublesome would-be axion is then
eaten by the new $U(1)^\prime$ gauge boson to give a massive $Z'$ at
the TeV scale.  An extra $U(1)^\prime$ gauge group can also be
motivated within the framework of grand unified theories (GUTs), arising 
as the relic of the breakdown of the unified gauge group. For
example, an $E_6$ GUT symmetry can be broken to the rank--5 subgroup
$SU(3)_C\times SU(2)_L\times U(1)_Y\times U(1)'$ where in general
$U(1)'=U(1)_{\chi} \cos\theta+U(1)_{\psi} \sin\theta$ \cite{E6}, and
the two anomaly-free $U(1)_{\psi}$ and $U(1)_{\chi}$ symmetries
originate from the breakings $E_6\to$ $SO(10)\times U(1)_{\psi}$,
$SO(10)\to$ $SU(5)\times$ $U(1)_{\chi}$. For a review see
\cite{U(1)-review} and for a discussion of the latest Tevatron and 
early LHC $Z'$ mass limits see \cite{Accomando:2010fz}.

With additional Abelian gauge symmetries it is also important to
ensure the cancellation of anomalies. This fits very nicely into
the framework of an $E_6$ GUT since, for any $U(1)^\prime$ that is a
subgroup of $E_6$, anomalies are cancelled automatically if the low
energy spectrum constitutes a complete $27$-plet.

Furthermore, within the class of $E_6$ models, there is a unique choice
of Abelian gauge group that allows zero charges for right-handed
neutrinos and thus large Majorana masses and a high scale see-saw
mechanism. This is the $U(1)_{N}$ gauge symmetry given by
$\theta=\arctan\sqrt{15}$ which is naturally achieved by GUT scale Higgses 
which develop VEVs in the ``right-handed neutrino'' 
component. The choice of  $U(1)_{N}$ gauge group coupled
with complete $27$-plets of matter at low energy defines the
E$_6$SSM \cite{King:2005jy}. 

The right-handed neutrinos acquire heavy Majorana masses and may play
a role in the early Universe by
decaying unequally into final states with lepton number $L=\pm 1$, 
creating a cosmological lepton asymmetry.  Because the Yukawa
couplings of the new exotic particles of the model are not
constrained by the neutrino oscillation data, substantial values of
CP--violating lepton asymmetries can be induced even for a relatively
small mass of the lightest right--handed neutrino ($M_1 \sim
10^6\,\mbox{GeV}$) so that successful thermal leptogenesis may be
achieved without encountering any gravitino problem
\cite{King:2008qb}.

The extra $U(1)_{N}$ gauge symmetry survives to low energies and
forbids a bilinear term $\mu {H}_d {H}_u$ in the superpotential but
allows the interaction $\lambda S H_d H_u$. At the electroweak (EW)
scale, the scalar component of the SM singlet superfield $S$ acquires
a non-zero VEV, $\langle S \rangle=s/\sqrt{2}$, breaking $U(1)_N$ and
yielding an effective $\mu=\lambda s/\sqrt{2}$ term.  Thus the $\mu$
problem in the E$_6$SSM is solved in a similar way to the
next-to-minimal supersymmetric standard model (NMSSM)~\cite{nmssm}, but
without the accompanying problems of singlet tadpoles or domain
walls. 

Recently we discussed a constrained version of the E$_6$SSM
(cE$_6$SSM), based on a universal high energy soft scalar mass $m_0$,
soft trilinear coupling $A_0$ and soft gaugino mass $M_{1/2}$
\cite{Athron:2009ue,Athron:2009bs,Athron:2008np}.  We proposed a number of 
benchmark points, and calculated the SUSY and exotic spectrum which 
we found to have the following characteristics:
\begin{itemize}
\item a spin-1 $Z_N'$ gauge boson of mass around 1-2 TeV; 
\item light gauginos including a light gluino of mass $\sim M_3$ 
(typically 350-650 GeV), a light
wino-like neutralino and chargino pair of mass $\sim M_2$ (typically 100-200 GeV), and a light
bino-like neutralino of mass $\sim M_1$ (typically 60-120 GeV), where $M_i$ are the low
energy gaugino masses, which are typically driven small compared to the 
effective $\mu$ parameter (typically 700-1400 GeV) by
renormalisation group (RG) running;  
\item heavier sfermions (typically 800-1600 GeV), except for the lightest stop which may be 500-800 GeV;
\item possibly light
exotic colour triplet charge $1/3$ $D$-fermions, with masses
controlled by independent Yukawa couplings enabling them to be as light as 
the Tevatron limit of about 300 GeV.
\end{itemize}

In this paper, motivated by the light spectrum above, 
we consider it urgent and timely to discuss two of the most characteristic and striking LHC signatures of the 
cE$_6$SSM in considerably 
more detail than was done in \cite{Athron:2009ue,Athron:2009bs}.
Firstly, we discuss the $U(1)_N$ spin-1 $Z'$ gauge boson (referred to as $Z_N'$) and show how
it may decay into exotic states, including the exotic $D$-fermions and singlinos.
This increases its width compared to that for SM fermion decays only,  
making its line shape more easily observed. Secondly, 
we calculate the LHC production cross section of exotic $D$-fermions and discuss their decay patterns.  
We illustrate these features by considering two of the benchmark points previously proposed
in some detail. Crucially, we also give the numerical Feynman rules which will
enable further studies (e.g. by experimentalists) to be performed.

We note that the phenomenology of $D$-fermions has also been discussed the general 
framework of $E_6$ models in \cite{Kang:2007ib}, but not specifically for the 
cE$_6$SSM which provides a more predictive framework via the use of benchmark points.

The layout of the rest of the paper is as follows.
In Section 2 we review the cE$_6$SSM.
Section 3 discusses the LHC predictions of the cE$_6$SSM illustrated through two 
benchmark points.   
Section 4 concludes the paper. We then have one Appendix, where the numerical 
Feynman rules utilised in this work are given.

\section{The Constrained E$_6$SSM}

The E$_6$SSM is a supersymmetric model with three generations of complete $27$ multiplets 
of matter and a low energy gauge group
 $SU(3)_C\times SU(2)_L\times U(1)_Y\times U(1)_N$, where the $U(1)_N$ is 
specified by the charges given in Tab. \ref{charges} and the combination $U(1)_{\chi}
\cos\theta+U(1)_{\psi} \sin\theta$, with $\theta=\arctan\sqrt{15}$. 

  The $27_i$ of $E_6$, each containing a quark and lepton family, decompose
under the $SU(5)\times U(1)_{N}$ subgroup of $E_6$ as follows:
\be
27_i\to \ds\left(10,\,\ds{1}\right)_i+\left(5^{*},\,\ds{2}\right)_i
+\left(5^{*},\,-\ds{3}\right)_i +\ds\left(5,-\ds{2}\right)_i
+\left(1,\ds{5}\right)_i+\left(1,0\right)_i\,.
\label{4}
\ee
The first and second quantities in the brackets are the $SU(5)$
representation and extra $U(1)_{N}$ charge while $i$ is a family index
that runs from 1 to 3. From Eq.~(\ref{4}) we see that, in order to
cancel anomalies, the low energy (TeV scale) spectrum must contain
three extra copies of $5^*+5$ of $SU(5)$ in addition to the three
quark and lepton families in $5^*+10$. To be precise, the ordinary SM
families which contain the doublets of left-handed quarks $Q_i$ and
leptons $L_i$, right-handed up- and down-quarks ($u^c_i$ and $d^c_i$)
as well as right-handed charged leptons, are assigned to
$\left(10,\ds{1}\right)_i+\left(5^{*},\,\ds{2}\right)_i$.
Right-handed neutrinos $N^c_i$ should be associated with the last term
in Eq.~(\ref{4}), $\left(1,0\right)_i$.  The next-to-last term in
Eq.~(\ref{4}), $\left(1,\ds{5}\right)_i$, represents SM-singlet
fields $S_i$ which carry non-zero $U(1)_{N}$ charges and therefore
survive down to the EW scale.  The three pairs of $SU(2)$-doublets
($H^d_{i}$ and $H^u_{i}$) that are contained in
$\left(5^{*},\,-\ds{3}\right)_i$ and $\left(5,-\ds{2}\right)_i$ have
the quantum numbers of Higgs doublets, and we shall identify one of
these pairs with the usual MSSM Higgs doublets, with the other two
pairs being ``inert'' Higgs doublets which do not get VEVs. The other
components of these $SU(5)$ multiplets form colour triplets of exotic
fermions $D_i$ and $\overline{D_i}$ with electric charges $-1/3$ and
$+1/3$ respectively. The matter content and correctly normalised
Abelian charge assignment are in Tab.~\ref{charges}.

\begin{table}[ht]
  \centering
  \begin{tabular}{|c|c|c|c|c|c|c|c|c|c|c|c|c|c|}
    \hline
 & $Q$ & $u^c$ & $d^c$ & $L$ & $e^c$ & $N^c$ & $S$ & $H_2$ & $H_1$ & $D$ &
 $\overline{D}$ & $H'$ & $\overline{H'}$ \\
 \hline
$\sqrt{\frac{5}{3}}Q^{Y}_i$
 & $\frac{1}{6}$ & $-\frac{2}{3}$ & $\frac{1}{3}$ & $-\frac{1}{2}$
& $1$ & $0$ & $0$ & $\frac{1}{2}$ & $-\frac{1}{2}$ & $-\frac{1}{3}$ &
 $\frac{1}{3}$ & $-\frac{1}{2}$ & $\frac{1}{2}$ \\
 \hline
$\sqrt{{40}}Q^{N}_i$
 & $1$ & $1$ & $2$ & $2$ & $1$ & $0$ & $5$ & $-2$ & $-3$ & $-2$ &
 $-3$ & $2$ & $-2$ \\
 \hline
  \end{tabular}
  \caption{\it\small The $U(1)_Y$ and $U(1)_{N}$ charges of matter fields in the
    E$_6$SSM, where $Q^{N}_i$ and $Q^{Y}_i$ are here defined with the correct
$E_6$ normalisation factor required for the RG analysis.}
  \label{charges}
\end{table}
If there are only complete matter multiplets at low energy, the gauge
couplings do not unify in a single step. Therefore one can either
proceed with two-step unification, leading to unification at the
string scale \cite{Howl:2007zi} or add incomplete multiplets.

In this paper we follow the latter path and require a further pair of
superfields $H'$ and $\overline{H}'$ with a mass term $\mu'
{H'}{\overline{H}'}$ from incomplete extra $27'$ and $\overline{27'}$
representations surviving to low energies. Anomaly cancellation is still
guaranteed since $H'$ and $\overline{H}'$ originate from the $27'$ and
$\overline{27'}$ supermultiplets.  Previous analysis reveals that the
unification of the gauge couplings in the E$_6$SSM can be achieved for
any phenomenologically acceptable value of $\alpha_3(M_Z)$, consistent
with the measured low energy central value, unlike in the MSSM which
requires significantly higher values of $\alpha_3(M_Z)$, well above
the central measured one \cite{unif-e6ssm}.

The superpotential of the E$_6$SSM contains many Yukawa couplings,
including interactions between the SM singlets, $S_i$ to both the
three generations of Higgs-like fields and the new exotic $D$-fermion
fields, as well as interactions between the exotic $D$-fermions and
inert Higgs fields with ordinary matter (the first two generations of the Higgs-like
fields), which are new in comparison to the SM.   Since some of these new
interactions violate baryon number conservation and induce
non-diagonal flavour transitions there should be some symmetry structure 
suppressing or forbidding the dangerous terms.  A structure to do this can 
arise from a family symmetry at the GUT scale \cite{Howl:2008xz}. 

In the scenarios considered in this paper, following previous work 
\cite{King:2005jy, Athron:2009ue,Athron:2009bs},  to suppress baryon number violating
and flavour changing processes we postulate a $Z^{H}_2$ symmetry
under which all superfields except one pair of $H^d_{i}$ and $H^u_{i}$
(say $H_d\equiv H^d_{3}$ and $H_u\equiv H^u_{3}$) and one SM-type
singlet field ($S\equiv S_3$) are odd.  The $Z^{H}_2$ symmetry reduces
the number of the Yukawa interactions, and together with a further
assumed hierarchical structure of the Yukawa interactions, we can
simplify the form of the E$_6$SSM superpotential
substantially. Keeping only Yukawa interactions whose couplings are
allowed to be of order unity leaves us with the following phenomenologically viable superpotential,
\beq
\ba{rcl}
W_{\rm E_6SSM}&\simeq &\lambda S (H_{d} H_{u})+\lambda_{\alpha}
S(H^d_{\alpha} H^u_{\alpha})+ \kappa_i S (D_i\overline{D}_i)\\[2mm]
&&+h_t(H_{u}Q)t^c+h_b(H_{d}Q)b^c+ h_{\tau}(H_{d}L)\tau^c+
\mu'(H^{'}\overline{H^{'}}),
\ea
\label{cessm8}
\eeq
where $\alpha,\beta=1,2$ and $i=1,2,3$, and where the superfields
$L=L_3$, $Q=Q_3$, $t^c=u^c_3$, $b^c=d^c_3$ and $\tau^c=e^c_3$ belong
to the third generation and $\lambda_i$, $\kappa_i$ are dimensionless
Yukawa couplings with $\lambda \equiv \lambda_3$.  Since the
right-handed neutrino has no charge under the $U(1)_N$ gauge symmetry,
nor under the SM gauge group, we assume that all right--handed
neutrinos are relatively heavy so that they can be integrated out. The
$SU(2)_L$ doublets $H_u$ and $H_d$, and singlet $S$ which are even
under the $Z^{H}_2$ symmetry, now play the role of Higgs fields,
generating the masses through EWSB, while the other generations of
these Higgs like fields remain inert.  The $H_u$ and $H_d$ fields
provide masses to the up-type and down-type quarks and leptons
respectively, just as in the MSSM, while $S$, which must acquire a
large VEV to induce sufficiently large masses for the $Z_N'$ boson,
also give masses to the exotic $D$-fermions and inert Higgs bosons
from Yukawa interactions, $\lambda_{\alpha} S(H^d_{\alpha}
H^u_{\alpha})$ and $\kappa_i S (D_i\overline{D}_i)$ . The couplings
$\lambda_i$ and $\kappa_i$ should be large enough to ensure the exotic
fermions are sufficiently heavy to avoid conflict with direct particle
searches at present and past accelerators. One generation of the new
Yukawa couplings (chosen to be the 3rd generation) should also be
large enough so that the evolution of the soft scalar mass $m_S^2$ of
the singlet field $S$ results in negative values of $m_S^2$ at low
energies, triggering the breakdown of the $U(1)_{N}$ symmetry.

However the $Z^{H}_2$ can only be approximate since under an exact
$Z^{H}_2$ decays of the exotic particles would be forbidden.
Therefore, while Eq.~\ref{cessm8} does not induce any proton decay,
some suppressed couplings can, and so to prevent rapid proton decay in
the E$_6$SSM we should still introduce a discrete symmetry to
play the role of $R$--parity in the MSSM. We give two examples of
possible symmetries that can achieve that.

 If $H^d_{i}$, $H^u_{i}$, $S_i$, $D_i$, $\overline{D}_i$ and the quark
superfields ($Q_i$, $u^c_i$, $d^c_i$) are even under a discrete
$Z^L_2$ symmetry while the lepton superfields ($L_i$, $e^c_i$,
$N^c_i$) are odd (Model I) then the allowed superpotential is
invariant with respect to a $U(1)_B$ global symmetry. The exotic
$\overline{D_i}$ and $D_i$ are then identified as diquark and
anti-diquark, i.e. $B_{D}=-2/3$ and $B_{\overline{D}}=2/3$. An
alternative possibility is to assume that the exotic quarks $D_i$ and
$\overline{D_i}$ as well as lepton superfields are all odd under
$Z^B_2$ whereas the others remain even. In this case (Model II) the
$\overline{D_i}$ and $D_i$ are leptoquarks \cite{King:2005jy}.  With
both of these symmetries the MSSM particle content behaves like it
does under $R$--parity, with the subset of particles present in the
standard model and Higgs (and also inert Higgs) bosons being even
under this generalised $R$--parity, while their supersymmetric
partners are odd and therefore, as usual, must be pair produced, and
upon decaying will always give rise to a stable lightest
supersymmetric particle (LSP).  However the exotic $D$-fermions are
odd and so must be pair produced and will decay into an LSP, while
their scalar superpartners are even and can be singly produced.

After $U(1)_N$ and EW symmetry breaking the Higgs fields, $H_u$, $H_d$
and $S$ give a physical Higgs spectrum of three CP--even, one CP-odd
and two charged states.  Two of the CP--even Higgs bosons tend to be
rather heavy, with one mass being close to the $Z'$ boson mass
$M_{Z'}$ and the other almost degenerate with the CP--odd Higgs boson and
the charged Higgs states.  The remaining CP--even Higgs bosons is
always light irrespective of the SUSY breaking scale, and has an upper
bound on its mass, as in the MSSM and NMSSM, but in the E$_6$SSM it
can be heavier than $110-120\,\mbox{GeV}$ even at tree level. In the
two--loop approximation the lightest Higgs boson mass does not exceed
$150-155\,\mbox{GeV}$ \cite{King:2005jy,Accomando:2006ga}. However 
for the benchmarks considered in the constrained model defined below
\cite{Athron:2009ue,Athron:2009bs} the lightest Higgs mass was in the
range $115-121\,\mbox{GeV}$, and the points we selected for the study
in this paper have light Higgs masses just above the LEP bound.

While the simplified superpotential of the E$_6$SSM in
Eq.~\ref{cessm8} only has six more couplings than the MSSM
superpotential, the soft breakdown of SUSY gives rise to many new
parameters.  The number of fundamental parameters can be reduced
drastically though within a constrained version of the model.
Constrained SUSY models imply that all soft scalar masses are set to
be equal to $m_0$ at some high energy scale $M_X$, taken here to be
equal to the GUT scale, all gaugino masses $M_i(M_X)$ are equal to
$M_{1/2}$ and trilinear scalar couplings are such that
$A_i(M_X)=A_0$. Thus the cE$_6$SSM is characterised by the following
set of Yukawa couplings, which are allowed to be of the order of
unity, and universal soft SUSY breaking terms,
\begin{equation}
\lambda_i(M_X),\quad \kappa_i(M_X),\quad h_t(M_X),\quad h_b(M_X), \quad 
h_{\tau}(M_X), \quad m_0, \quad M_{1/2},\quad A_0,
\label{3}
\end{equation}
where $h_t(M_X)$, $h_b(M_X)$ and $h_{\tau}(M_X)$ are the usual
$t$--quark, $b$--quark and $\tau$--lepton Yukawa couplings, and
$\lambda_i(M_X)$, $\kappa_i(M_X)$ are the extra Yukawa couplings
defined in Eq.~(\ref{cessm8}). The universal soft scalar and trilinear
masses correspond to an assumed high energy soft SUSY breaking
potential of the universal form,
\begin{equation}
V_{soft}= m_0^227_i27_i^*+A_0Y_{ijk}27_i27_j27_k +h.c.,
\end{equation}
where $Y_{ijk}$ are generic Yukawa couplings from the trilinear terms
in Eq.~(\ref{cessm8}) and the $27_i$ represent generic fields from
Eq.~(\ref{4}), and in particular those which appear in
Eq.~(\ref{cessm8}). In previous analyses
\cite{Athron:2009ue,Athron:2009bs} we always set  $m_0^2$ positive 
for correct EWSB and to simplify the analysis assume that all parameters in
Eq.~(\ref{3}) are real and $M_{1/2}$ is positive.  The set of
cE$_6$SSM parameters in Eq.~(\ref{3}) should in principle be
supplemented by $\mu'$ and the associated bilinear scalar coupling
$B'$. However, since $\mu'$ is not constrained by the EWSB and the
term $\mu'H'\overline{H}'$ in the superpotential is not suppressed by
$E_6$, the parameter $\mu'$ was assumed to be $\sim 10\,\mbox{TeV}$ so
that $H'$ and $\overline{H}'$ decoupled from the rest of the particle
spectrum. As a consequence the parameters $B'$ and $\mu'$ are
irrelevant for the analysis \cite{Athron:2009ue,Athron:2009bs}.

In addition several of the parameters specified above are fixed by
experimental measurements and the RG flow.  This means that the
particle spectrum and many phenomenological aspects of the model can be
determined from only eight free parameters, which in previous analyses
have been taken to be\footnote{Note that $m_0$, $M_{1/2}$ and $A_0$
have been replaced by $v$, $\tan \beta$ and $s$ through the EWSB
conditions, in as similar manner to the way $|\mu|$ and $B$ are traded 
for $\tan\beta$ and $v$ in the MSSM.} $\{ \lambda_i$, $\kappa_i$, $s$, $\tan \beta$\}, 
which can be compared to the cMSSM with $\{m_0, M_{1/2}, A, \tan \beta,
sign(\mu)$\}, and could be reduced further by considering scenarios with
some Yuakawa coupling universality or other well motivated relations
between the Yukawa couplings at the GUT scale.

To calculate the particle spectrum within the cE$_6$SSM a private
spectrum generator has been written, based on some routines and the class 
structure of SOFTSUSY 2.0.5 \cite{Allanach:2001kg} and employing two-loop RG
equations (RGEs) for the gauge and Yukawa couplings together with
two-loop RGEs for $M_a(Q)$ and $A_i(Q)$ as well as one-loop RGEs for
$m_i^2(Q)$, where $Q$ is the renormalisation scale. The details of the
procedure we followed, including the RGEs for the E$_6$SSM and the
experimental and theoretical constraints can be found in
\cite{Athron:2009ue,Athron:2009bs}.

\section{LHC signatures of the cE$_6$SSM}

\subsection{Benchmark spectra and couplings}

In previous publications we presented a set of  ``early
discovery'' benchmark points which should be discovered using first
LHC data \cite{Athron:2009ue} and a set of slightly heavier
(``late discovery'') benchmarks
\cite{Athron:2009bs} to illustrate the wider range of possible cE$_6$SSM
scenarios which could be discovered at the LHC. Here we select two of
these points for a more detailed phenomenological study, focussing on
the $Z_N^\prime$ and the new exotic colored states.  For this we have
chosen the ``early discovery'' benchmark C (BMC) and a heavier,
qualitatively different benchmark 4 (BM4).  The mass spectra for these are
given in Tab.~\ref{table:benchmarks}.
\begin{table}[tbp]
\begin{center}
\begin{tabular}{|c|c|c|}
\hline  &                                     \textbf{\footnotesize BMC} & \textbf{\footnotesize BM4} \\\hline
\footnotesize $\tan \beta$                    &\footnotesize 10      &\footnotesize 30            \\[-1.5mm]
\footnotesize $\lambda_3(M_X)$                &\footnotesize -0.378  &\footnotesize -0.38         \\[-1.5mm]
\footnotesize $\lambda_{1,2}(M_X)$            &\footnotesize 0.1     &\footnotesize 0.1           \\[-1.5mm]
\footnotesize $\kappa_3(M_X)$                 &\footnotesize 0.42    &\footnotesize 0.16          \\[-1.5mm]
\footnotesize $\kappa_{1,2}(M_X)$             &\footnotesize 0.06    &\footnotesize 0.16          \\[-1.5mm]
\footnotesize $s$[TeV]                        &\footnotesize 2.7     &\footnotesize 5.0           \\[-1.5mm]
\footnotesize $M_{1/2}$[GeV]                  &\footnotesize 388     &\footnotesize 725           \\[-1.5mm]
\footnotesize $m_0$ [GeV]                     &\footnotesize 681     &\footnotesize 1074          \\[-1.5mm]
\footnotesize $A_0$[GeV]                      &\footnotesize 645     &\footnotesize 1726          \\
\hline
\footnotesize $m_{\tilde{D}_{1}}(3)$[GeV]     &\footnotesize 1465    &\footnotesize 312           \\[-1.5mm]
\footnotesize $m_{\tilde{D}_{2}}(3)$[GeV]     &\footnotesize 2086    &\footnotesize 2623          \\[-1.5mm]
\footnotesize $\mu_D(3)$[GeV]                 &\footnotesize 1747    &\footnotesize 1621          \\[-1.5mm]
\footnotesize $m_{\tilde{D}_{1}}(1,2)$[GeV]   &\footnotesize 520     &\footnotesize 312           \\[-1.5mm]
\footnotesize $m_{\tilde{D}_{2}}(1,2)$[GeV]   &\footnotesize 906     &\footnotesize 2623          \\[-1.5mm]
\footnotesize $\mu_D(1,2)$[GeV]               &\footnotesize 300     &\footnotesize 1621          \\
\hline
\footnotesize $|m_{\chi^0_6}|$[GeV]           &\footnotesize 1054    &\footnotesize 1950          \\[-1.5mm]
\footnotesize $m_{h_3}\simeq M_{Z'}$[GeV]     &\footnotesize 1021    &\footnotesize 1889          \\[-1.5mm]
\footnotesize $|m_{\chi^0_5}|$[GeV]           &\footnotesize 992     &\footnotesize 1832          \\
\hline
\footnotesize $m_S(1,2)$[GeV]                 &\footnotesize 1001    &\footnotesize 1732          \\[-1.5mm]
\footnotesize $m_{H_2}(1,2)$[GeV]             &\footnotesize 627     &\footnotesize 1117          \\[-1.5mm]
\footnotesize $m_{H_1}(1,2)$[GeV]             &\footnotesize 459     &\footnotesize 220           \\[-2mm]
\footnotesize $\mu_{\tilde{H}}(1,2)$[GeV]     &\footnotesize 233     &\footnotesize 491           \\
\hline
\footnotesize $m_{\tilde{u}_1}(1,2)$[GeV]     &\footnotesize 911     &\footnotesize 1557          \\[-1.5mm]
\footnotesize $m_{\tilde{d}_1}(1,2)$[GeV]     &\footnotesize 929     &\footnotesize 1595          \\[-1.5mm]
\footnotesize $m_{\tilde{u}_2}(1,2)$[GeV]     &\footnotesize 929     &\footnotesize 1595          \\[-1.5mm]
\footnotesize $m_{\tilde{d}_2}(1,2)$[GeV]     &\footnotesize 964     &\footnotesize 1664          \\[-1.5mm]
\footnotesize $m_{\tilde{e}_2}(1,2,3)$[GeV]   &\footnotesize 849     &\footnotesize 1427          \\[-1.5mm]
\footnotesize $m_{\tilde{e}_1}(1,2,3)$[GeV]   &\footnotesize 765     &\footnotesize 1254          \\[-1.5mm]
\footnotesize $m_{\tilde{\tau}_2}$[GeV]       &\footnotesize 845     &\footnotesize 1363          \\[-1.5mm]
\footnotesize $m_{\tilde{\tau}_1}$[GeV]       &\footnotesize 757     &\footnotesize 1102          \\[-1.5mm]
\footnotesize $m_{\tilde{b}_2}$[GeV]          &\footnotesize 955     &\footnotesize 1491          \\[-1.5mm]
\footnotesize $m_{\tilde{b}_1}$[GeV]          &\footnotesize 777     &\footnotesize 1193          \\[-1.5mm]
\footnotesize $m_{\tilde{t}_2}$[GeV]          &\footnotesize 829     &\footnotesize 1248          \\[-1.5mm]
\footnotesize $m_{\tilde{t}_1}$[GeV]          &\footnotesize 546     &\footnotesize 837           \\
\hline
\footnotesize $|m_{\chi^0_3}|\simeq |m_{\chi^0_4}|\simeq
|m_{\chi^{\pm}_2}|$[GeV]
                                              &\footnotesize 674     &\footnotesize  1343         \\ [-1.5mm]
\footnotesize $m_{h_2}\simeq m_A 
\simeq m_{H^{\pm}}$[GeV]                      &\footnotesize 963     &\footnotesize 998           \\[-1.5mm]
\footnotesize $m_{h_1}$[GeV]                  &\footnotesize 115     &\footnotesize 114           \\ \hline
\footnotesize $m_{\tilde{g}}$[GeV]            &\footnotesize 353     &\footnotesize 642           \\[-1.5mm]
\footnotesize $|m_{\chi^{\pm}_1}|\simeq |m_{\chi^0_2}|$[GeV]
                                              &\footnotesize 109     &\footnotesize 206           \\[-1.5mm]
\footnotesize $|m_{\chi^0_1}|$[GeV]           &\footnotesize 61      &\footnotesize 116           \\
\hline
\end{tabular}
\caption{Parameters for
the ``early discovery'' benchmark point C (left) (from \cite{Athron:2009ue}) and ``late discovery''
benchmark point 4 (from \cite{Athron:2009bs}).}
\label{table:benchmarks}
\end{center}
\end{table}

These spectra both exhibit the characteristic cE$_6$SSM signature of a
heavy sfermion sector, with light gauginos.  Previously we observed
that in the cE$_6$SSM $m_0 \gtrsim M_{1/2}$ for all phenomenologically
viable points \cite{Athron:2009ue,Athron:2009bs}.  Additionally we
discovered that the low energy gluino mass parameter $M_3$ is driven
to be smaller than $M_{1/2}$ by RG running, due to the much larger
(super)field content of the E$_6$SSM in comparison to the MSSM (three
27's instead of three 16's).  This implies that the low energy gaugino
masses are all less than $M_{1/2}$ in the cE$_6$SSM, being given by
roughly\footnote{ These should be compared to the corresponding
  low energy values in the cMSSM (or mSUGRA), $M_3 \sim 2.7M_{1/2}$, $M_2 \sim
  0.8M_{1/2}$, $M_1 \sim 0.4M_{1/2}$.  Thus a particular value of $M_{1/2}$ in the 
  cE$_6$SSM gives the same gluino mass as a corresponding value 
  of $M_{1/2}$ in the cMSSM (or mSUGRA) approximately four times smaller. 
By contrast a particular value of $m_0$ in the cE$_6$SSM gives the same squark mass
as a corresponding value of $m_0$ 
in the cMSSM (or mSUGRA) very roughly of order one and a half times larger.
Thus as an extremely crude approximation 
$(m_0,M_{1/2})_{cE_6SSM} \rightarrow ((3/2)m_0,(1/4)M_{1/2})_{cMSSM}$ 
which underlines the cE$_6$SSM prediction of relatively heavy squarks and relatively light gluinos. 
Note that this is the least sensitive region of the recent cMSSM analysis by CMS
\cite{Khachatryan:2011tk} and ATLAS \cite{Collaboration:2011hh}
.}
  $M_3 \sim 0.7M_{1/2}$, $M_2 \sim 0.25M_{1/2}$, $M_1 \sim
0.15M_{1/2}$. These two features imply that
the sfermions of ordinary matter will always be heavier than the
lightest gauginos, and the lightest SUSY states will include of a
light gluino of mass $\sim M_3$, a light wino-like neutralino and
chargino pair of mass $\sim M_2$, and a light bino-like neutralino of
mass $\sim M_1$. 

%%Notice also that the enlarged field content means that at one--loop 
%%order the QCD beta function (accidentally) vanishes in the E$_6$SSM, 
%%and at two loops it loses asymptotic freedom (though the gauge couplings 
%%remain perturbative at high energy).

The heavier spectrum of BM4 is due to a significantly larger choice
for the singlet vacuum expectation value, $s=\sqrt{2}\langle S \rangle = 5$ TeV
as opposed to $s = 2.7$ TeV in BMC. While
substantial variation in the spectra can be produced by varying the
new Yukawa couplings associated with exotic interactions, $\langle S
\rangle$ is linked to the spectrum through the 
EWSB conditions and $U(1)$$_N$ $D$-terms, so choosing a particular
value places restrictions on the masses and in general the larger
$\langle S \rangle$ the heavier the spectrum.

The $U(1)_N$ gauge coupling, $g_1^\prime$, is fixed by gauge
coupling unification with the RG flow leading to $g_1^\prime(M_Z)
\approx g_1(M_Z)$.  This means that $\langle S \rangle$ fixes the mass
of the $Z^\prime_N$, since $M_{Z^\prime} \sim g_1^\prime \langle
S\rangle$ and this leads to $M_{Z^\prime} = 1890$ GeV for BM4 and
$M_{Z^\prime} = 1021$ GeV for BMC, affecting the discovery potential
at the LHC, as will be discussed later.

Another consequence of this is that the couplings to the $Z_N^\prime$
are also highly constrained in this model since they are given by the
gauge coupling and the $U(1)_N$ charges.  Variation of these couplings
between benchmark points comes only from mass mixing of gauge
eigenstates, the scale dependence of $g_1^\prime$, and two-loop
running effects.  This variation can be seen in Appendix
\ref{appendix:FeynRules} where the $Z_N^\prime$ Feynman rules are
presented for our two benchmarks.  However, despite this, there is still
considerable room for different phenomenologies for a given
$M_{Z^\prime}$ (or equivalently $\langle S \rangle$), and this can
also strongly impact on the Drell Yan production cross section of the
$Z_N^\prime$.
   
 For example the exotic colored fermions can be light or heavy, since
 their masses are given by $\mu_{D_i} = \frac{1}{\sqrt{2}}\kappa_i s$,
 and if $\kappa$ universality is not assumed\footnote{At least one
   $\kappa_i$ coupling must be large to generate EWSB.} it is possible
 to obtain two $\kappa_i(M_S)$ (where $M_S$ is the SUSY breaking
 scale) small enough that the exotic fermions are just above their
 mass limit ($300$ GeV), as BMC illustrates. 
However the masses of the scalar partners to the exotic coloured
fermions also have soft mass contributions which tend to increase with
$M_{Z^\prime}$, and as a result only one of the two scalars can be light,
and it is unlikely that both scalars will be available as $Z^\prime_N$
decay modes.  However, even without small $\kappa_i$, it is still
possible to have a light exotic sfermion due to large mixing, and this
is demonstrated in BM4.

The inert Higgsino masses, $\mu_{H_\alpha} =
\frac{1}{\sqrt{2}}\lambda_i s$, may also be light for a sufficiently
small $\lambda_\alpha$ coupling and this is the case in BMC (and to a
lesser extent BM4).  However it is possible to also have all
$\lambda_i$ large, giving Higgsinos of a TeV or above, and not
available for the $Z^\prime_N$ to decay into.   The scalar inert Higgs masses can
be very light depending on the particular parameters chosen.  However
as with the exotic sfermions, due to the soft mass
contribution, there is usually a hierachy between the inert Higgs
bosons of a particular generation.

Both the inert and exotic coloured states have large $U(1)_N$ charges
which mean they can play an important role in $Z^\prime_N$ phenomenology,
as well as also producing interesting signatures from direct
production.

All the sfermions of ordinary matter are rather heavy, with the sfermions 
in BM4 being substantially heavier than in BMC,  arising from the influence 
of the larger $M_{Z^\prime}$ in the EWSB conditions.  The stops tend to 
be the lightest of the sfermions and due to large mixing the lightest stop 
is the only ordinary sfermion which can be really light with the possibility 
of being just above $400$ GeV.  In the two benchmarks here we have
$m_{\tilde{t_1}}= 546$ GeV for BMC and $m_{\tilde{t_1}}= 837$~GeV for
BM4, due as usual to the heavier $M_{Z^\prime}$.

The light SUSY states that are always present in the spectrum include a
light gluino $\tilde g$, two light neutralinos $\chi_1^0,\chi_2^0$,
and a light chargino $\chi_1^\pm$. The lightest neutralino $\chi_1^0$
is essentially pure bino, while $\chi_2^0$ and $\chi_1^\pm$ are the
degenerate components of the wino. Since these particles are composed
primarily from states that do not couple to the $Z_N^\prime$, they do not
play a large role in the $Z_N^\prime$ phenomenology.
In addition there are other neutralinos $\chi_3^0$ and $\chi_4^0$
which are essentially pure Higgsino states and 
$\chi_5^0$ and $\chi_6^0$ associated with the third family singlino
$\tilde{S}$ and the $\tilde{Z'}$ gaugino.

Finally there are also two light inert singlinos not shown 
explicitly in Table~\ref{table:benchmarks} (the SUSY partners to the 
two families of inert singlet scalars $S(1,2)$) whose masses
 are given by suppressed couplings that are assumed to be small
 enough so that they do not perturb the RG running of the other
 couplings. So, although these masses are not precisely fixed in
 previous analyses of the cE$_6$SSM spectrum, they are assumed to be
 very light.  These particles then guarantee that there will be a
 substantial non-SM contribution to the $Z_N^\prime$ width. However, if
 there is also other light exotic matter, then it can also make a
 significant contribution,  as will be discussed later in the
 paper.

\subsection{Tevatron and LEP limits}
The presence of light states (neutralinos, chargino and inert singlinos) 
in the E$_6$SSM particle spectrum raises serious concerns that they could have
already been observed at the Tevatron and/or even earlier at LEP. For example, 
the light neutralino and chargino states could be produced at the 
Tevatron \cite{Baer:1992dc}. Recently, the CDF and D0 collaborations set 
a stringent lower bound on chargino masses using searches for SUSY with 
a trilepton final state \cite{trilepton}. These 
searches ruled out chargino masses below $164\,\mbox{GeV}$. However this 
lower bound on the chargino mass was obtained by assuming that the 
corresponding chargino and neutralino states decay predominantly into the 
LSP and a pair of leptons. In our case, the lightest neutralino 
and chargino states are expected to decay via virtual $Z$ and $W$ exchange, 
and then predominantly into the LSP and a pair of quarks. As a 
consequence the lower limit on the mass of charginos that is set by the 
Tevatron is not directly applicable to the benchmark scenarios that we 
consider here. Instead in our study we use the $95\%\,\mbox{C.L.}$ 
lower limit on the chargino mass of about $100\,\mbox{GeV}$ that was 
set by LEP II \cite{Kraan:2005vy}.

LEP experiments also set stringent constraints on the masses and couplings 
of neutral particles that interact with the $Z$ boson.
Since inert singlinos have masses below $M_Z/2$, the $Z$ boson could decay
into these states. However the couplings of these exotic states to the
$Z$--boson are rather small due to their singlino nature \cite{Hall:2010ix}.
Consequently their contribution to the $Z$ boson decay width and
the corresponding branching ratios are negligible. Due to the
small $Z$ couplings, the production of light inert singlinos at 
LEP was extremely suppressed, which allowed these states to escape 
detection at LEP. 

Nevertheless the presence of light inert singlinos could lead to 
other phenomena which could be observed at LEP. In the case of BMC,
$\chi_{1}^0 \chi_{1}^0$ and $\chi_{1}^0 \chi_{2}^0$ could be produced
followed by their decay into
inert singlino via virtual $Z$ exchange, resulting in $q\bar{q}q'\bar{q}'$
and missing energy in the final state. LEP has set limits on the cross 
section of  $e^{+}e^{-}\to\chi_2^0\chi_1^0\,(\chi_1^{+}\chi_1^{-})$
in the case where the subsequent decay is predominantly
$\chi_2^0\to q\bar{q}\chi^0_1$ $(\chi_1^{\pm}\to q\bar{q}'\chi^0_1)$  
\cite{Abbiendi:2003sc}. Unfortunately, these bounds are not 
directly applicable to our study, but they do demonstrate that
it was difficult to observe 
$$
e^{+}e^{-}\to X+Y\to q\bar{q}q'\bar{q}'+\Big/ \hspace{-0.3cm E_T}\,,
$$
where $X$ and $Y$ are neutral particles, if the corresponding production cross 
section was $0.1-0.3\,\mbox{pb}$. In the case of the BMC the lightest and
second lightest neutralinos have rather small couplings to the $Z$ boson. The 
corresponding relative couplings are of the order $(M_W/\mu)^2$.
Since the selectron is also heavy in the considered scenario the production 
cross sections of $\chi_{1}^0 \chi_{1}^0$ and $\chi_{1}^0 \chi_{2}^0$
are suppressed by $O(\frac{1}{M^4})$ where $M\sim 700-800\,\mbox{GeV}$. 
At LEP energies the cross sections of colourless particle
production through $s$-channel $\gamma/Z$ exchange are typically a few 
picobarns, so the production cross sections of $\chi_{1}^0 \chi_{1}^0$ 
and $\chi_{1}^0 \chi_{2}^0$ in the case of BMC are expected to be 
of the order of $10^{-2}\,\mbox{pb}$ or even less.
Thus BMC could not be ruled out by LEP experiments.

The Higgsino states are much heavier with the degenerate Higgsinos
$\chi_{3,4}^0$ and $\chi_2^\pm$ having masses given by $\mu = \lambda
s/\sqrt{2}$ in the range 675--830 GeV for both benchmark points considered.
The remaining neutralinos are dominantly third generation singlino
and the gaugino partner of the $Z_N^\prime$ with masses approximately
given by $M_{Z'}$.

The Higgs spectrum for all the benchmark points contains a very light
SM--like CP--even Higgs boson $h_1$ with a mass close to the LEP limit
of 115 GeV, making it accessible to LHC or even Tevatron.  The heavier
CP--even Higgs $h_2$, the CP--odd Higgs $A_0$, and the charged Higgs
$H^{\pm}$ are all closely degenerate with masses above
900 GeV making them difficult to discover.  The remaining mainly
singlet CP--even Higgs $h_3$ is closely degenerate with the $Z_N'$.

Tevatron, LEP and other experiments also set limits on the mass of the $Z_N'$ boson,
$Z-Z_N'$ mixing and masses of exotic scalars ($\tilde D$). The direct searches
at the Fermilab Tevatron $(p\overline{p}\to Z'_N\to l^{+}l^{-})$ exclude $Z_N'$
with mass below $892\,\mbox{GeV}$ \cite{Accomando:2010fz}
\footnote{Slightly weaker lower bound on the mass of the $Z_N'$ boson 
was obtained in \cite{Erler:2010uy}. Note that these bounds assume $Z_N'$ boson decays
only into quarks and leptons. If the width increases by about a factor of two due to exotics
and SUSY particles (as will be the case
for the benchmarks studied in this paper) then this would reduce the branching ratio into 
charged leptons also by a factor of two, which we estimate would reduce the
mass limit quoted in  \cite{Accomando:2010fz} from $892\,\mbox{GeV}$ down to about
$820\,\mbox{GeV}$.}. At the LHC, the $Z'$ boson that appears in the $E_6$ inspired models can
be discovered if it has a mass below $4-4.5\,\mbox{TeV}$
\cite{ZprimeE6}.  The determination of its couplings should be
possible if $M_{Z'}\lesssim 2-2.5\,\mbox{TeV}$ \cite{Dittmar:2003ir}.
The precision EW tests bound the $Z-Z'$ mixing angle to be around $(-1.5)-0.7\times 10^{-3}$
\cite{Erler:2009jh}. Recent results from Tevatron searches for dijet
resonances \cite{CDFtevDijet} rule out scalar diquarks with mass less than
$630$ GeV. However, scalar leptoquarks may be as light as $300$ GeV       
since at hadron colliders they are pair produced through gluon fusion \cite{Aktas:2005pr}.

\subsection{Phenomenology}

In this subsection, we focus on the phenomenology of the two benchmark 
points  (BMC and BM4), in order to illustrate two of the most striking 
cE$_6$SSM predictions: the exotic contributions 
to the heavy jet rate and the existence of a $Z_N'$ boson 
with an enhanced and resolvable width due to its additional 
decays into exotic states.

Tab.~\ref{table:widths} presents the cE$_6$SSM $Z_N'$ partial decay 
widths in all available channels. Apart from the leading SM decays
into quarks ($q$) and leptons ($l$),
one can notice, amongst the  cE$_6$SSM channels, the dominance of
the decay into singlinos (collectively denoted by $\tilde S$), 
whose mass we have set at 10 and 30 GeV, for the two generations, 
respectively\footnote{Notice that their contribution to the total 
$Z'$ width is typically always about 30\%, irrespectively of their actual 
mass, so long that the singlino masses remains within 
the boundaries established in \cite{Hall:2010ix},
as space effects are minimal for the considered $Z'$ masses.}. 
Next in line in order of importance are the exotic fermion (specifically, 
$D$-fermion, when open) and inert Higgsino ($\tilde H$) channels. 
The genuine SUSY contributions into gauginos ($\tilde \chi$) are 
never sizable while exotic scalars ($\tilde D$) 
and sfermions ($\tilde f$) count negligibly.
At times (here for BM4), decays into inert Higgs ($H^{0 / \pm}_{\alpha, i}$) 
states can also be tangible. Overall, non-SM contributions to the
c$E_6$SSM $Z_N'$ width are of order 100\% for both benchmarks considered. 
   \begin{table}[tbp]
\begin{center}
\begin{tabular}{|l|c|c|}
\hline  
 $Z_N'$ partial width [GeV]                                 & {\bf BMC}             & {\bf BM4} \\
\hline
 $\Gamma(Z_N'\to l^+l^-)$  ($l=e,\mu$ or $\tau$)          & 0.41              &  0.77   \\
 $\Sigma_l \Gamma(Z_N'\to \nu_{l}\overline{\nu}_{l})$   (all neutrinos)          & 0.87                &  1.64   \\

 $\Sigma_l \Gamma(Z_N'\to l^+l^-,\nu_{l}\overline{\nu}_{l})$                (all leptons)          & 2.10                &  3.96   \\
 $ \Sigma_q  \Gamma(Z_N'\to q\bar q)$               (all quarks)           & 5.31                & 10.08   \\
$\Sigma_{i} \Gamma(Z_N'\to D_i\bar D_i)$               (exotic fermions)  & 3.49                &  0.00   \\
 $\Sigma_{\alpha} \Gamma(Z_N'\to \tilde H_\alpha \tilde H_\alpha )$     (inert Higgsinos)  & 3.09                &  5.19   \\
 $\Sigma_{\alpha} \Gamma(Z_N'\to \tilde S_{\alpha} \tilde S_{\alpha})$     (singlinos)        & 4.05                &  7.63   \\
 $\Sigma_i  \Gamma(Z_N'\to \tilde D_i\tilde D_i)$      (exotic scalars)   & 0.00                &  0.19   \\
 $\Sigma_f \Gamma(Z_N'\to \tilde f \tilde f)$     (sfermions)        & 0.00                &  0.010   \\
 $\Sigma_{\alpha} \Gamma(Z_N'\to H_\alpha H_\alpha)$     (inert Higgses)    & 0.026               &  0.39   \\
  $\Sigma_j\Gamma(Z_N'\to \tilde\chi_j \tilde\chi_j)$ (gauginos)        & 6.50$\times10^{-4}$ &  7.92$\times10^{-5}$ \\
\hline
 $\Gamma_{\rm tot}$                    (all)                & 18.07               &  27.45  \\
\hline
\end{tabular}
\caption{$Z_N'$ widths for 
the ``early discovery'' benchmark point C (left) (from \cite{Athron:2009ue}) and ``late discovery''
benchmark point 4 (from \cite{Athron:2009bs}). The index $i$ is summed over three families,
the index $\alpha$ is summed over the two inert families of exotics while $j$ is summed over the light neutralino and chargino states. The leptonic branching ratio into $l^+l^-$ is given by $Br(l^+l^-)\approx 0.023$ for BMC and $Br(l^+l^-)\approx 0.028$ for BM4, as compared to the value calculated by ignoring the exotics and SUSY 
partners of $Br(l^+l^-)\approx 0.055$ in both cases. The Drell-Yan cross-section may be defined
in terms of two parameters $c_u$ and $c_d$ which are defined and discussed in \cite{Carena:2004xs}.
In the limit where exotics and SUSY partners are ignored their values for this model are given by 
 $c_u\approx 5.9\times 10^{-4}$ and $c_d\approx 1.5\times 10^{-3}$  \cite{Accomando:2010fz}. 
Since $c_u$ and $c_d$ 
are both proportional to $Br(l^+l^-)$ they will therefore be reduced for both benchmarks due to the
presence of exotics and SUSY partners by about a factor of two in each case.
For BMC we find $c_u\approx 2.4\times 10^{-4}$ and $c_d\approx 0.61\times 10^{-3}$, while for BM4 we find
$c_u\approx 3.0\times 10^{-4}$ and $c_d\approx 0.75\times 10^{-3}$.}
\label{table:widths}
\end{center}
\end{table}

The presence of light exotic particles and gauginos gives rise to non-standard decays of 
the $Z_N^\prime$ gauge boson. Indeed, exotic states, that originate from the $Z_N^\prime$ decays, 
subsequently decay resulting in the four-fermion final states with and without missing energy. 
For example, the $Z_N^\prime$ can decay into a pair of second lightest singlinos. Then second 
lightest singlino sequentially decays into the lightest one and a fermion--antifermion pair 
mainly via a virtual $Z$. Since lightest singlino is stable it leads to the missing energy 
in the final state. Because second lightest singlino tend to be relatively light it decays 
predominantly into light quarks and leptons. At the same time the decays of the $Z_N^\prime$ 
into $D$-fermions (or Inert Higgsinos) give rise to the final states that contain four third 
generation fermions and missing energy as will be clarified later. Because $Z_N^\prime$ is 
relatively heavy its decay products, which appear in the corresponding exotic final states, 
should have sufficiently high energies. Therefore some of them (in particular, charged leptons) 
might be observed at the LHC.

\subsubsection{Benchmark C}
We now discuss the details of 
the ``early discovery'' BMC in Table~\ref{table:benchmarks} corresponding to a lighter spectrum
first observable at the LHC with 7 TeV, then subsequently amenable to detailed study at 14 TeV. 
Since BMC leads to signatures that may be more readily discovered and studied at the LHC we 
discuss this benchmark in considerably more detail than the subsequent one, which is included 
mainly for comparison.

\subsubsection*{$Z_N'$ bosons}

Fig.~\ref{BMC_LHC_7} (top frame) shows
the differential distribution in invariant mass of the lepton pair $l^+l^-$ (for one species
of lepton $l=e,\mu$ or $\tau$) in Drell-Yan production at the LHC for $\sqrt s=7$ TeV, assuming 
a sequential $Z'$ (that is, with the same mass as 
in the cE$_6$SSM but with SM-like couplings, i.e.\ no additional matter) as well as 
a cE$_6$SSM 
$Z'$ field with and without light exotic quarks
and inert Higgsinos\footnote{We have 
three generations of the exotic quarks but only two of inert Higgs. For convenience,
in the legends of the plots we only refer to the former.
Also note that we always include the other width contributions, according to Tab.~\ref{table:widths}.}.

This distribution is promptly measurable even at the lower energy stage 
of the CERN collider with a high resolution and would 
enable one to not only confirm the existence of a $Z'$ state but also to
establish the possible presence and nature of additional exotic matter, by simply fitting
to the data the width of the $Z'$ resonance, its height at the resonance point
and its profile in the interference region with the SM channels ($\gamma$- and $Z$-mediated).  In fact,
for our choice of $\mu_{D_i}$, $\mu_{H_i}$ and $M_{Z'}$, the
$Z_N'$ total width varies from $\approx 7$ GeV (in case of SM-only matter) 
to $\approx 18$ GeV (in case of additional cE$_6$SSM matter). In particular, notice the different
normalisation around the $Z'$ resonance of the three curves in Fig.~\ref{BMC_LHC_7} (top frame)\footnote{Clearly, in order to perform such an
exercise, the $Z'$ couplings to ordinary matter ought to have been previously established elsewhere, as a
modification of the latter may well lead
to effects similar to those induced by the additional matter present in our model. (Recall that in our model
$Z_N'$ couplings to SM particles and exotic matter are simultaneously fixed.)}.

Another $Z'$ observable (alongside the cross section normalisation and its line shape 
near and below the $Z'$ peak) which will be useful to access $Z'$ couplings is 
the forward-backward asymmetry (here denoted by AFB$_{l^+\l^-}$).  
Fig.~\ref{BMC_LHC_7} (middle frame) indeed shows a sizable difference
in its shape (around the $Z'$ mass resonance, especially) between the 
cases of a sequential $Z'$ and a cE$_6$SSM $Z_N'$, albeit difficult to measure at the 7 TeV 
LHC (assuming 1 fb$^{-1}$ of total accumulated luminosity). Remarkably, the shape 
(and normalisation) of  AFB$_{l^+l^-}$ is essentially the same in the cE$_6$SSM 
irrespective of its particle content, so that the ability of accessing the $Z'$ couplings 
in such a model does not require a knowledge of its spectrum beforehand.

Fig.~\ref{BMC_LHC_14} (top and middle frame) reinstates
the above phenomelogical aspects at 14 TeV, with the added bonus of much 
larger event rates (by a factor of 6 or so around
the $Z'$ peak) and luminosity (which could be up 
to 300 fb$^{-1}$ at the end of the collider lifetime).

\subsubsection*{Exotics}

If exotic particles of the nature described here do exist at low scales, they could possibly 
be accessed through direct pair hadroproduction. However, as remarked in \cite{King:2005jy}, 
the corresponding fully inclusive and differential cross sections are 
sufficient only in the case of exotic $D$-fermions (because they
are pair produced via QCD interactions) while inert Higgsinos most likely remain 
inaccessible (as their pair production is induced by EW interactions). 

\begin{figure}[tbp]
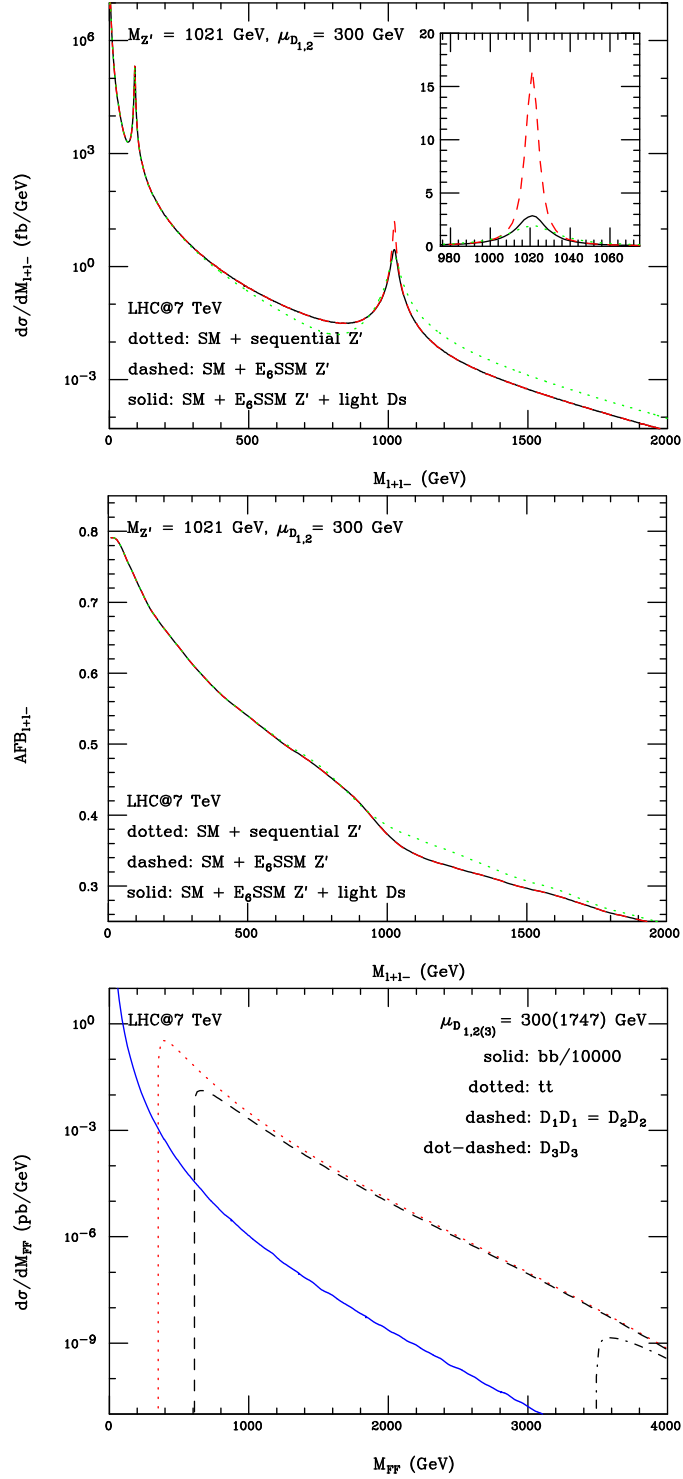

 \begin{center}
\resizebox{!}{6.5cm}
{\includegraphics[width=\linewidth,angle=90]{Benchmark-Fig7a.ps}}
\resizebox{!}{6.5cm}
{\includegraphics[width=\linewidth,angle=90]{Benchmark-Fig7b.ps}}
\resizebox{!}{6.5cm}
{\includegraphics[width=\linewidth,angle=90]{Benchmark-Fig8.ps}}
\caption{Results for benchmark C at the 7 TeV LHC. 
{\it Top:} Differential cross sections for Drell-Yan production, 
with respect to the lepton pair invariant mass. 
{\it Middle:} Forward-backward asymmetries. 
{\it Bottom:} Production cross sections of exotic $D$-fermion pairs, 
in comparison to bottom- and top-quark pair production. The total production 
rates are $\sigma(D_1D_1)=\sigma(D_2D_2)=3 \, {\rm pb}$ and
$\sigma(D_3D_3)=0.0005\, {\rm fb}$.
\label{BMC_LHC_7}
}
\end{center}
\end{figure}

\begin{figure}[tbp]
 \begin{center}
\resizebox{!}{6.5cm}
{\includegraphics[width=\linewidth,angle=90]{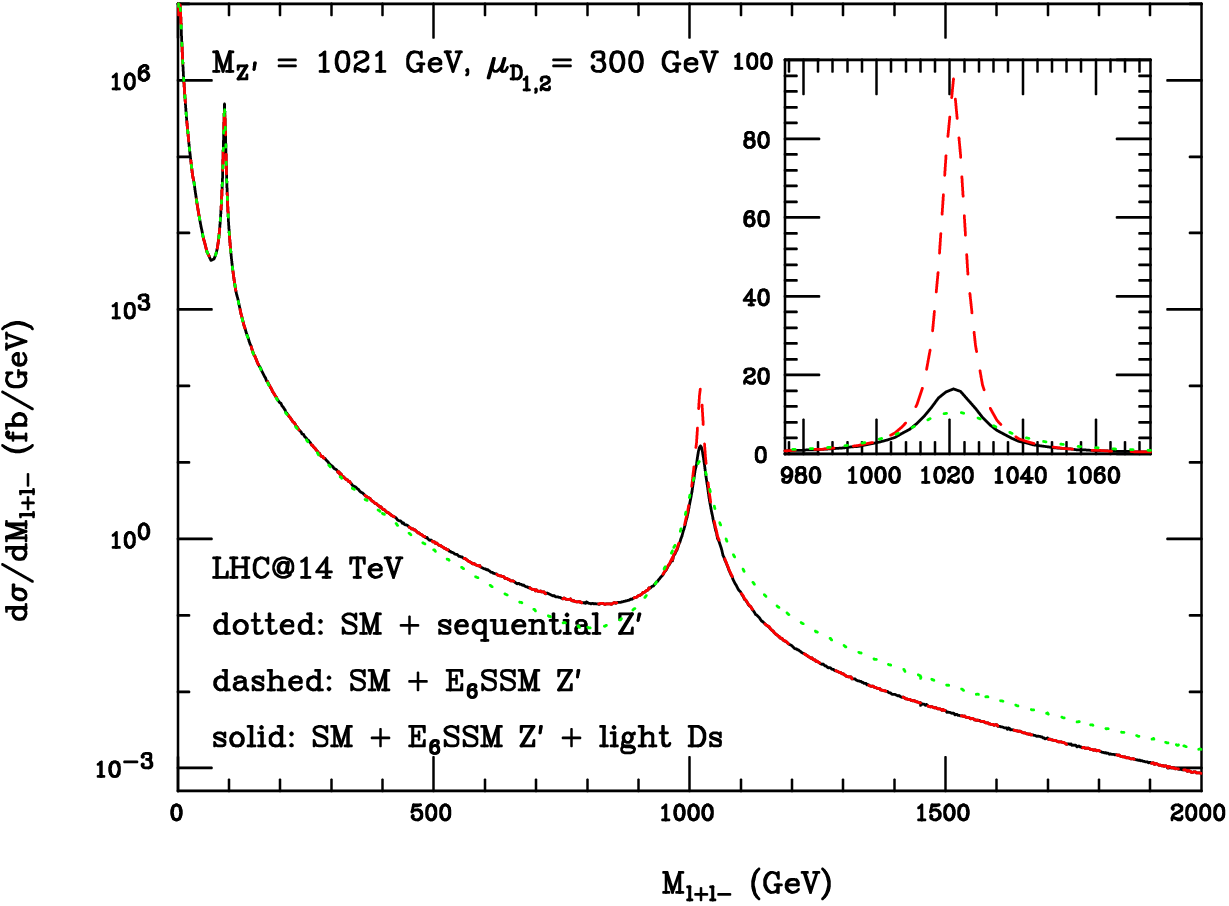}}
\resizebox{!}{6.5cm}
{\includegraphics[width=\linewidth,angle=90]{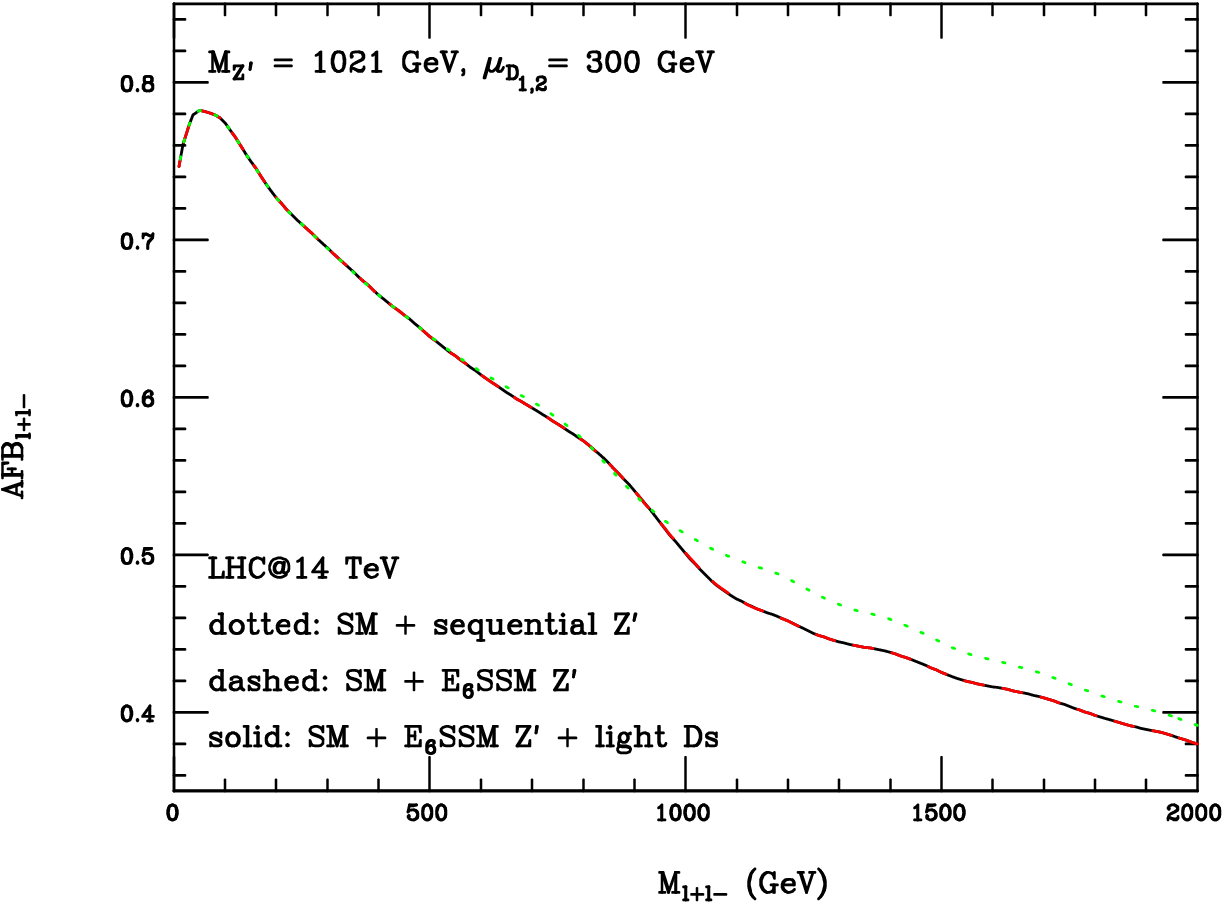}}
\resizebox{!}{6.5cm}
{\includegraphics[width=\linewidth,angle=90]{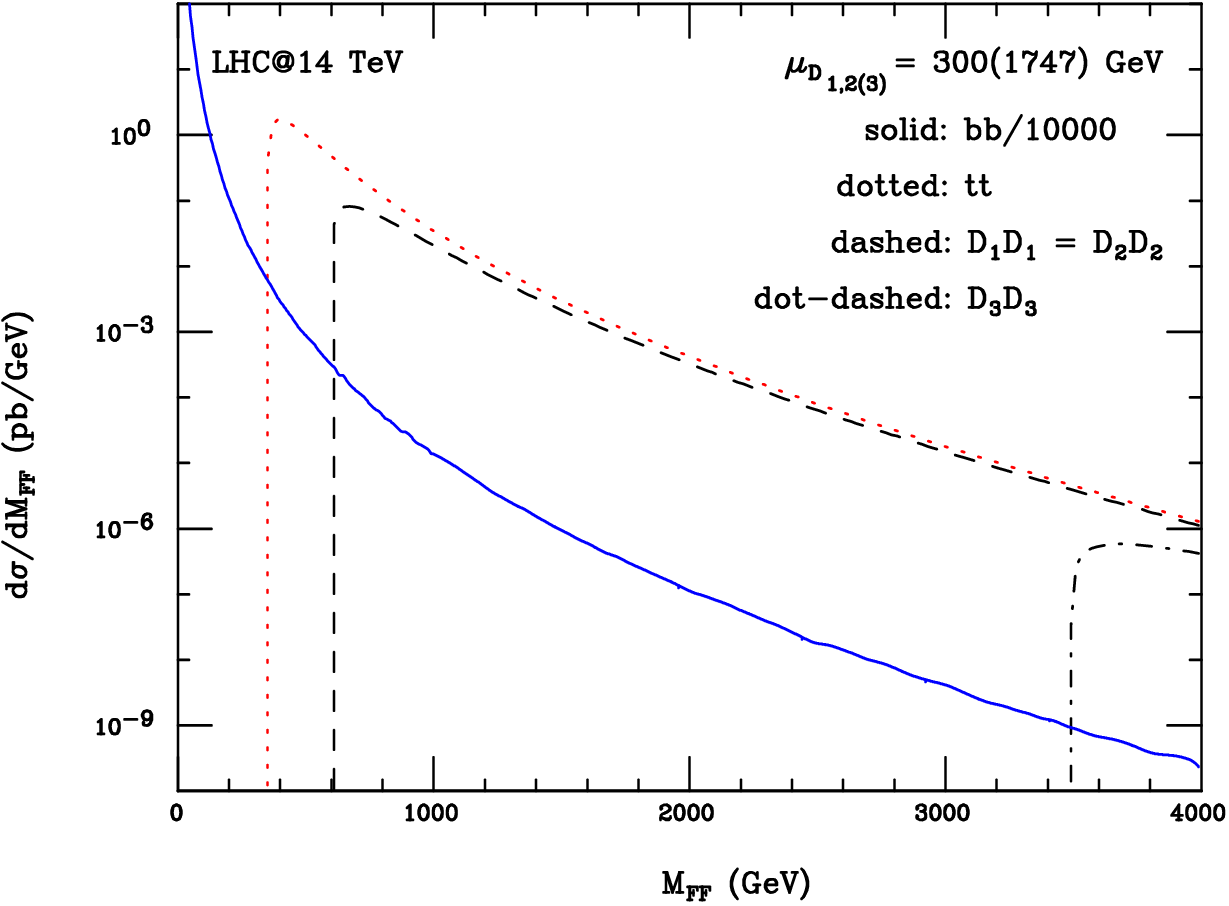}}
\caption{Results for benchmark C at the 14 TeV LHC. 
{\it Top:} Differential cross sections for Drell-Yan production, 
with respect to the lepton pair invariant mass. 
{\it Middle:} Forward-backward asymmetries. 
{\it Bottom:} Production cross sections of exotic $D$-fermion 
pairs, in comparison to bottom- and top-quark pair production. 
The total production rates are 
$\sigma(D_1D_1)=\sigma(D_2D_2)=25 \, {\rm pb}$ and
$\sigma(D_3D_3)=0.5\, {\rm fb}$.
 \label{BMC_LHC_14} }
\end{center}
\end{figure}

Therefore, we plot  the production
cross section of exotic $D$-fermion pairs, in comparison to those for bottom-
and top-quark pair production, in the bottom frame of both 
Figs.~\ref{BMC_LHC_7} and \ref{BMC_LHC_14}, for an LHC with
7 and 14 TeV centre-of-mass energy respectively, using CTEQ5L with $Q^2=\hat s$. 
Although the detectable final states 
resulting from exotic $D$-fermion production do depend on the underlying nature 
of the exotic particles, we find that experimental signatures involve
multi-jet states containing identifiable $b$-hadrons, whether
produced via $t$-resonances or not, as we shall now discuss.

As outlined in \cite{King:2005jy}, the lifetime and decay modes of the exotic 
$D$-fermions are determined by the operators that break the $Z_2^{H}$ symmetry. 
When $Z_2^H$ is broken significantly exotic fermions can produce a remarkable 
signature\footnote{If $Z_2^{H}$ is only slightly broken exotic quarks may live for a long time, 
and form compound states with ordinary quarks. This means that at future colliders it may be
possible to study the spectroscopy of new composite scalar leptons or baryons. Also one can observe 
quasi-stable charged colourless fermions with zero lepton number.}. Since, according to our initial 
assumptions, the $Z_2^{H}$ symmetry is mostly broken by operators involving quarks and leptons
of the third generation, the exotic $D$-fermions decay either via
$$
\ba{llll}
\overline{D}\to t+\tilde{b}\,,\qquad & \overline{D}\to b+\tilde{t}\, , \qquad & D \to  \overline{t}+\tilde{b}^*\,,\qquad & D\to\overline {b}+\tilde{t}^*,
\ea
$$
if exotic $\overline{D}_i$ fermions are diquarks or via
$$
\ba{llll}
D\to t+\tilde{\tau}\,, \qquad & D\to \tau^- +\tilde{t}\,, \qquad & \overline{D}\to \bar{t}+\tilde{\tau}^*\,, \qquad & \overline{D}\to \tau^+ +\tilde{t}^*\, , \\
D\to b+\tilde{\nu}_{\tau}\, , \qquad & D\to \nu_{\tau}+\tilde{b}\,,\qquad & \overline{D}\to \bar{b}+\tilde{\nu}^*_{\tau}\, , \qquad & \overline{D}\to \nu_{\tau}+\tilde{b}^* \\
\ea
$$
if exotic $D$-fermions are leptoquarks. In general, sfermions decay into the
corresponding fermion and a neutralino, so one expects that each diquark will decay
into $t$- and $b$-quarks while a leptoquark will produce a $t$-quark and $\tau$-lepton
in the final state with rather high probability. Thus the presence of light exotic 
$D$-fermions in the particle spectrum could result in an appreciable enhancement of 
the cross section of either $pp\to t\overline{t}b\overline{b}+X$ and $pp\to b\overline{b}b\overline{b}+X$
if exotic $D$-fermions are diquarks or $pp\to t\overline{t}\tau^+{\tau^-}+X$ and
consequently $pp\to b\overline{b}\tau^+{\tau^-}+X$ if $D$-fermions are leptoquarks\footnote{It 
is worth to remind the reader here that the production cross sections
of $pp\to t\overline{t}b\overline{b}+X$ and $pp\to t\overline{t}\tau^+{\tau^-}+X$
in the SM are suppressed {\sl at least} by a factor $\left(\ds\frac{\alpha_s}{\pi}\right)^2$ and
$\left(\ds\frac{\alpha_W}{\pi}\right)^2$, respectively, as compared
to the cross section of $t\overline{t}$ pair production (and, similarly, for $t$-quarks replaced by 
$b$-quarks).}.

Each $t$-quark decays
into a $b$-quark while a $\tau$-lepton gives one charged lepton $l$ in the final 
state with a probability of $35\%$. Therefore both these scenarios would ultimately 
generate an excess in the $b$-quark production cross section. Thus the presence of exotic 
$D$-fermions alters the SM data samples involving $t\overline{t}$ production and decay as well 
as direct $b\overline{b}$ production. 

A detailed LHC analysis will be required to establish the feasibility of extracting 
this excess due to the light exotic $D$-fermions predicted by our model. However, our 
results clearly show that, for the discussed parameter configuration, the position is 
favourable, as the product of production rates and branching ratios for these channels are typically 
larger than the expected four-body SM cross sections involving heavy quarks. 

 \begin{figure}[tbp]
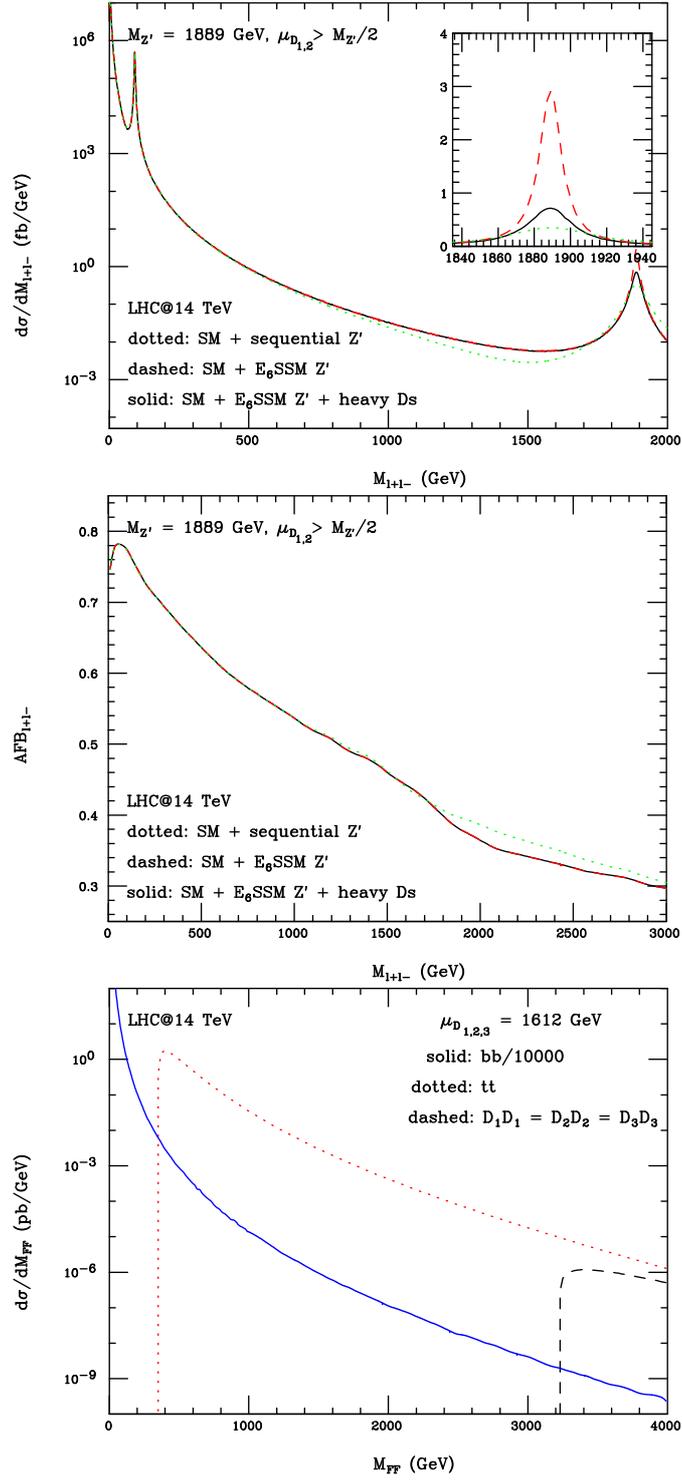

 \begin{center}
\resizebox{!}{6.5cm}
{\includegraphics[width=\linewidth,angle=90]{Benchmark-Fig13a.ps}}
\resizebox{!}{6.5cm}
{\includegraphics[width=\linewidth,angle=90]{Benchmark-Fig13b.ps}}
\resizebox{!}{6.5cm}
{\includegraphics[width=\linewidth,angle=90]{Benchmark-Fig14.ps}}
\caption{Results for benchmark 4 at the 14 TeV LHC. 
{\it Top:} Differential cross sections for Drell-Yan production, with respect 
to the lepton pair invariant mass. {\it Middle:} Forward-backward asymmetries. 
{\it Bottom:} Production cross sections of exotic $D$-quark pairs, in comparison 
to bottom- and top-quark pair production. The total production rates are 
$\sigma(D_1D_1)=\sigma(D_2D_2)=\sigma(D_3D_3)=0.9\, {\rm fb}$.
 \label{BM4_LHC_14} }
\end{center}
\end{figure}

\subsubsection{Benchmark point 4}

Having discussed in detail the phenomenology of the ``early discovery''
BMC, we can be relatively brief when discussing
BM4 in Table~\ref{table:benchmarks}, which represents the case of a heavier spectrum for 
the cE$_6$SSM, not necessarily discoverable at 7 TeV, hence dubbed ``late discovery''. 
BM4 is included mainly for comparison with BMC in order to fairly show that the cE$_6$SSM does 
not always lead to such a light spectrum.

\vspace{-2mm}
\subsubsection*{$Z_N'$ bosons}

For this parameter configuration, with a rather heavy
$Z_N'$, cross sections are much smaller, beyond detectability at the 7 TeV LHC.
We therefore only present results for the higher energy stage of the CERN collider, 
in Fig.~\ref{BM4_LHC_14} (top and middle frame) for the $Z'$ line shape and forward-backward 
asymmetry. The pattern that emerges here is very much in line with 
that of the previous benchmark, albeit with reduced production rates overall. However,
the $Z_N'$ should remain detectable at the 14 TeV LHC after full luminosity is collected 
(Also note that the absolute value of the corrections to the $Z_N'$ width due to cE$_6$SSM 
particles is somewhat larger here, growing by about 13 GeV.). 

\vspace{-2mm}
\subsubsection*{Exotics}

In contrast, the exotic $D$-fermions are much heavier for this benchmark and their detectability, 
even at 14 TeV, remains debatable. This is illustrated in Fig.~\ref{BM4_LHC_14} (bottom frame), 
where their inclusive cross section is only of order 10 fb (including all three generations) 
and requires very high invariant mass final states, where the control of the SM background is uncertain.  

Nevertheless the particle spectrum of BM4 contains relatively light exotic scalars ($\tilde{D}$).
Because these exotic scalars have masses about  $300\,\mbox{GeV}$ they are expected to be leptoquarks.
So light leptoquarks should be efficiently produced at the LHC. They decay into quark--lepton final
states mainly through $Z_2^H$ violating operators involving quarks and leptons of the third generation,
i.e. $\tilde{D} \rightarrow t \tau$. This leads to an enhancement of $pp \rightarrow t \bar t \tau \bar{\tau}$
(without missing energy) at the LHC.

\section{Conclusions}
We have previously proposed a constrained version of the
Exceptional Supersymmetric Standard model, the cE$_6$SSM, based on a
universal high energy soft scalar mass $m_0$, soft trilinear mass
$A_0$ and soft gaugino mass $M_{1/2}$. The cE$_6$SSM
predicts a characteristic SUSY spectrum containing a light gluino, a
light wino-like neutralino and chargino pair, and a light bino-like
neutralino, with other sparticle masses except the lighter stop being
much heavier. In addition, cE$_6$SSM allows the possibility of
light exotic colour triplet charge $1/3$ $D$ fermions and scalars,
and predicts an observable $Z_N'$ spin-1 gauge boson.

In this paper, motivated by the fact that the cE$_6$SSM allows the 
spectrum above to be quite light and observable with the first data from the LHC, 
we have focussed on two of the most characteristic and striking LHC signatures of the 
cE$_6$SSM, namely the prediction of a $Z_N'$ gauge 
boson and exotic $D$-fermions, and the interplay between these two predictions.
In particular we have shown how the 
$Z_N'$ gauge boson may decay into exotic $D$-fermions, increasing its width and 
modifying its line shape. For example, we find that the width may increase by a factor of two, which effectively 
reduces the Drell-Yan cross-section into charged lepton pairs also by a factor of two,
relaxing the current Tevatron limits from 892 GeV down to about 820 GeV.
In addition we have calculated the LHC production cross section of the $D$-fermions
and discussed their decay patterns.  

The added value of the cE$_6$SSM, compared to previous studies, is that it provides
a predictive framework for the experimental study of such signatures via the use of benchmark points. 
We illustrated this by considering two of the benchmark points previously proposed
in some detail. The first benchmark point C, which has low values of
$(m_0,M_{1/2})$ around $(700,400)$~GeV and a $Z_N'$ gauge boson with mass 
around 1 TeV, gave rise to signatures corresponding to an ``early LHC discovery'' using
``first data''. 
We also examined benchmark point 4 with higher values of
$(m_0,M_{1/2})$ around $(1100,700)$~GeV and a $Z'$ gauge boson with mass 
around 2 TeV, providing
a more challenging scenario corresponding to late discovery using all accumulated data at the CERN collider.
Note that these values of $(m_0,M_{1/2})$ in the cE$_6$SSM yield a squark and gluino spectrum
roughly equivalent to that in the cMSSM with $m_0$ about $3/2$ times larger and $M_{1/2}$ about $4$ times smaller
than the corresponding cE$_6$SSM values.

If a $Z_N'$ gauge boson and $D$-fermions were discovered at the LHC, 
identified by measurements of their mass, cross-section and decay signatures as discussed here,
this would
not only represent a revolution in particle physics, but would also
point towards a possible underlying high energy E$_6$ gauge structure, providing
the first glimpse into superstring theory.

\vspace{-4mm}
\section*{Acknowledgements}
\vspace{-4mm} P.A.~would like to thank M.~Sch\"onherr, D.~St\"ockinger, H.~St\"ockinger-Kim 
and A.~Voigt for helpful comments. S.F.K. thanks P.A. for hospitality during a visit to Dresden.
R.N. would like to thank V.~Barger, E.~E.~Boos, M.~Drees, J.~P.~Kumar and X.~R.~Tata for fruitful discussions.
The work of S.M. is partially supported by the NExT Institute. 
S.F.K. and S.M. acknowledge partial support from the STFC Rolling Grant ST/G000557/1.
D.J.M. acknowledges partial support from the STFC Rolling Grant ST/G00059X/1.
The work of R.N. was supported by the U.S. Department of Energy under Contract DE-FG02-04ER41291.

\newpage

\begin{appendix}

\section{Feynman rules}
\label{appendix:FeynRules}
In this appendix the $Z^\prime_N$ Feynman rules of the E$_6$SSM for the considered benchmarks are presented.  
\begin{figure}[H]
 \begin{center}%%\resizebox{!}{10cm}
\resizebox{!}{10cm}
{\includegraphics{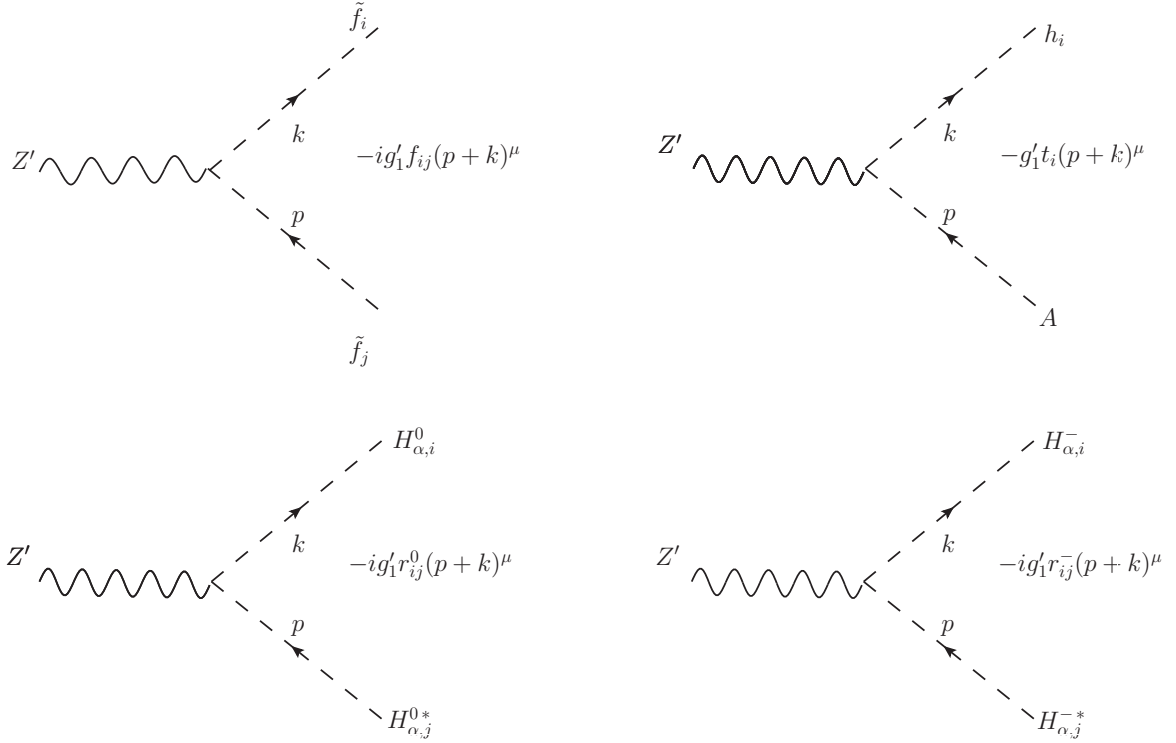}}
\end{center}
\caption{ Feynman rules: Z$^\prime_N$ coupling to scalars.}
\label{Fig:FeynScal}
\end{figure}

The couplings shown in Fig.~\ref{Fig:FeynScal} are determined as follows.  For the scalar partners of 
fermions with substantial mixing the couplings are given by
\begin{eqnarray}
f_{11} &=& (\tilde{Q}^N_{f_L}  \cos^2{\theta_{\tilde{f}}} - \tilde{Q}^N_{f^c} \sin^2{\theta_{\tilde{f}}}),
\nonumber\\ f_{22} &=& (\tilde{Q}^N_{f_L} \sin^2{\theta_{\tilde{f}}} - \tilde{Q}^N_{f^c}
\cos^2{\theta_{\tilde{f}}}), \nonumber\\ f_{12}=f_{21} &=& -(\tilde{Q}^N_{f^c}+\tilde{Q}^N_L)
\sin{\theta_{\tilde{f}}} \cos{\theta_{\tilde{f}}},
\end{eqnarray}
 where $\theta_{\tilde{f}}$ is the mixing 
\BE
\VL \sfe \\ \sfz \VR = \ML \costf & \sintf \\ -\sintf & \costf \MR 
                       \VL \sfl \\ \sfr \VR 
\label{Scalar_rotation}
\EE

 %% The couplings are then \BE \bar{\chi}^{0}_i Z_\mu i 0.5g_1^\prime\gamma^\mu \Phi_{ii}^*( -g_A^{ij}\gamma^5)\Phi_{jj} \chi^{0}_j, \EE where no sum over indices is implied for $\Phi_{ii}^*$ and $\Phi_{jj}$.

\noindent The relation between the Higgs gauge and mass eigenstates is $H_i^0 = U_{ji}^{-1}H_j + i V_{ji}^{-1}A_j$, where $H_i^0 = \{H_u^0,H_d^0,S\}$ and $A_j = \{A, G^\prime, G^0\}$, where the form of $V$ can be read off from Eqs.~(58)-(59) of Ref.~\cite{King:2005jy} and $U$ is found when the CP-even Higgs mass matrix is diagonalised.   The Higgs $Z^\prime_N$ Feynman rules shown in Fig.~\ref{Fig:FeynScal} then take the form $t_j = Q_i^\prime U_{ji}^{-1}V_{1i}^{-1} (p +k)^\mu$.  %the matrix which relates the CP-odd eigenstates to the %%The $V^{-1}_{ji}$ are given by $V^{-1}_{ji}A_j =  p_i/\sqrt{2} A + q_i/\sqrt{2} G' + r_i/\sqrt{2} G_0$, where $p_i = ( c_{\beta} c_{\psi}, s_\beta c_\psi, s_\psi)$ and $q_i = (-c_\beta s_\psi, -s_\beta s_\psi, c_\psi)$ and $r_i = (-s_\beta, c_\beta, 0)$ in the basis $(H_u, H_d, S)$.  The $U_{ij}$ are found by diagonalising the CP-even mass matrix.

The inert Higgs come from two generations of 'up' and 'down' type doublets 
and each generation has 8 degrees of freedom. The charged and neutral components are almost degenerate, but are 
split by D-term contributions.  The physical states are formed by mixing the 'up' and 'down' type Higgs as follows:\begin{eqnarray} 
H_{\alpha, 1}^0 &=& \cos\theta^0_\alpha H_{\alpha}^{d, 0} + \sin\theta^0_\alpha H_{\alpha}^{u, 0}, \\
H_{\alpha, 2}^0 &=& \cos\theta^0_\alpha H_{\alpha}^{u, 0} - \sin\theta^0_\alpha H_{\alpha}^{d, 0}, \\
H_{\alpha, 1}^- &=& \cos\theta^-_\alpha H_{\alpha}^{d, -} + \sin\theta^-_\alpha H_{\alpha}^{u, +*}, \\
H_{\alpha, 2}^- &=& \cos\theta^-_\alpha H_{\alpha}^{u, + *} - \sin\theta^-_\alpha H_{\alpha}^{d, -}.
\end{eqnarray} and the couplings shown in Fig.~\ref{Fig:FeynScal} are then of the form \begin{eqnarray}
r^0_{11} &=& (\tilde{Q}^N_{H_1}  \cos^2{\theta^0}-\tilde{Q}^N_{H_2} \sin^2{\theta^0}), \;\;\; r^-_{11} = (\tilde{Q}^N_{H_1}  \cos^2{\theta^-}-\tilde{Q}^N_{H_2} \sin^2{\theta^-})\\ 
r^0_{22} &=& (\tilde{Q}^N_{H_1} \sin^2{\theta^0} - \tilde{Q}^N_{H_2}\cos^2{\theta^0}),\;\;\; r^-_{22} = (\tilde{Q}^N_{H_1} \sin^2{\theta^-} - \tilde{Q}^N_{H_2}\cos^2{\theta^-})  \\ 
r^0_{12} &=& (\tilde{Q}^N_{H_2}+\tilde{Q}^N_{H_1})\sin{\theta^0} \cos{\theta^0},  \;\;\; r^-_{12} = (\tilde{Q}^N_{H_2}+\tilde{Q}^N_{H_1})\sin{\theta^-} \cos{\theta^-}.
\end{eqnarray}

The numerical values of the scalar couplings for our benchmarks are given in Tab. \ref{table:sZpCoup}.

\begin{table}[h!]
\begin{center}
\begin{tabular}{|c|c|c|}
\hline  &                \textbf{BMC }   & \textbf{BM4}  \\\hline
Stops $g_1^\prime f_{11}$     &   $-0.0.04537$   &  $-0.05128 $  \\
Stops  $g_1^\prime f_{22}$  &  $0.05827$ & $-0.06423$ \\
Stops  $g_1^\prime f_{12}$    &   $0.04518$   &  $ -0.04682$   \\
\hline
Sbottoms $g_1^\prime f_{11}$   &   $-0.07910$   &  $-0.07624$   \\
Sbottoms $g_1^\prime f_{22}$&  $-0.1592$   &   $-0.1571$  \\
Sbottoms $g_1^\prime f_{12}$ &  $-0.01574$   &   $-0.03302$  \\
\hline
Staus $g_1^\prime f_{11}$  &  $0.09303$   &   $0.09377
$   \\
Staus $g_1^\prime f_{22}$  &  $0.1474$   &   $-0.1487$   \\
\hline
Sups    $g_1^\prime f_{11}$  &  $-0.06540$   &   $-0.06787$   \\
Sups $g_1^\prime f_{22}$  &  $0.07812$   &   $0.080821$   \\
\hline
Sdowns    $g_1^\prime f_{11}$  &  $0.07812$   &   $0.0821$   \\
Sdowns  $g_1^\prime f_{22}$  &  $-0.1562$   &   $-0.1616$   \\
\hline
Selectron $g_1^\prime f_{11}$  &  $0.09306$   &   $0.09377$   \\
Selectron $g_1^\prime f_{22}$  &  $-0.14737$   &   $-0.1487$   \\
\hline                
Scalar exotic D's 3rd Gen $g_1^\prime f_{11}$     &   $0.06668$   &  $ 0.04380
$  \\
Scalar exotic D's 3rd Gen   $g_1^\prime f_{22}$  & $0.001274$ & $0.02407$ \\
Scalar exotic D's 3rd Gen  $g_1^\prime f_{12}$    &   $0.1859$   &  $ 0.1953$   \\
\hline                
Scalar exotic D's 1st/2nd Gen $g_1^\prime f_{11}$     &   $0.04114$   &  $  0.04380$  \\
Scalar exotic D's 1st/2nd Gen   $g_1^\prime f_{22}$  & $0.002426$ & $0.02407$ \\
Scalar exotic D's 1st/2nd Gen  $g_1^\prime f_{12}$    &   $0.1888$   &  $ 0.1953 $   \\
\hline
Neutral inert Higgs  $g_1^\prime r^0_{11}$     &   $-0.1581$   &  $-0.1042 $  \\
Neutral inert Higgs   $g_1^\prime r^0_{22}$  & $0.06511$ & $0.01042$ \\
Neutral inert Higgs  $g_1^\prime r^0_{12}$    &   $0.1585$   &  $ 0.1870$   \\
Charged inert Higgs  $g_1^\prime r^-_{11}$     &   $-0.1443$   &  $-0.1013 $  \\
Charged inert Higgs   $g_1^\prime r^-_{22}$  & $0.05129$ & $0.007607$ \\
Charged inert Higgs  $g_1^\prime r^-_{12}$    &   $0.1674$   &  $0.1878$   \\
\hline
Higgs $g_1^\prime t_1$  & $-0.01964$  & $-0.006459$ \\
Higgs  $g_1^\prime t_2$  & $0.1183$  & $ 0.1210$ \\
 Higgs  $g_1^\prime t_3$  & $0.002314$  & $ 0.0001778$ \\
\hline
\end{tabular}
\caption{Scalar couplings to $Z^\prime_N$.}
\label{table:sZpCoup}
\end{center}
\end{table}

\begin{figure}[H]
 \begin{center}%%\resizebox{!}{10cm}
\resizebox{!}{10cm}
{\includegraphics{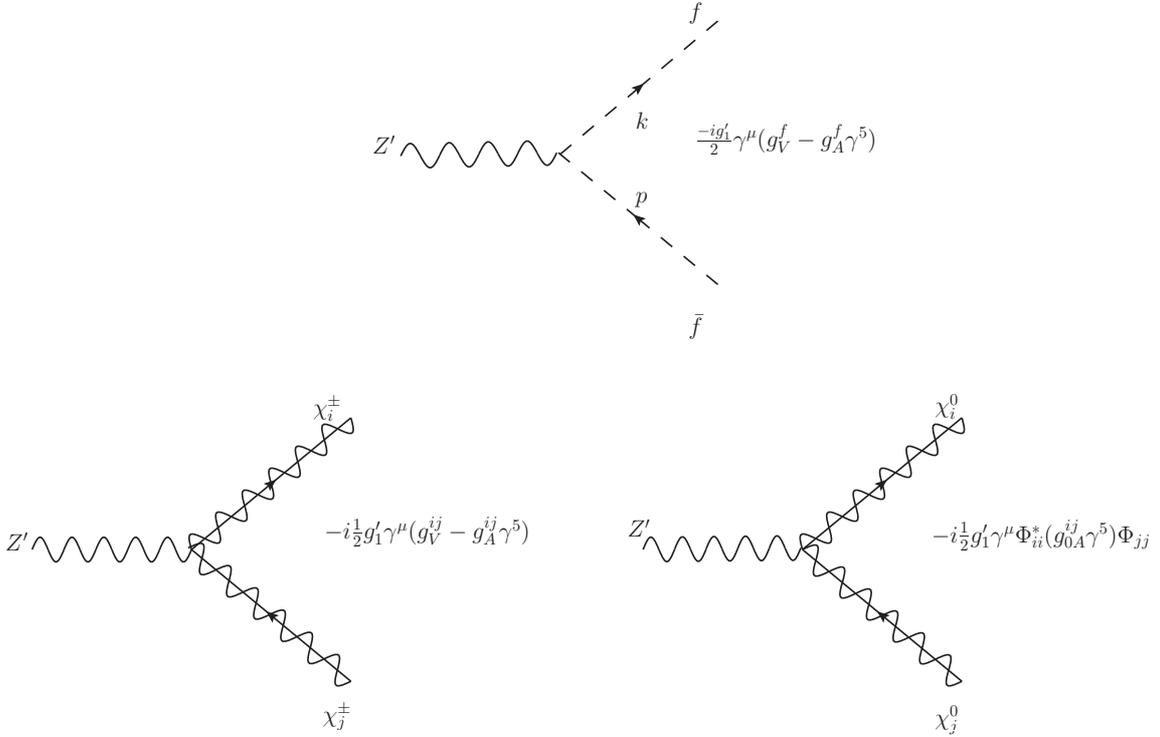}}
\end{center}
\caption{Feynman rules: $Z'$ couplings to fermions.}
\label{Fig:FeynFerm}
\end{figure}

The fermion Feynman rules are shown in Fig.~\ref{Fig:FeynFerm}.    The charginos masses are found by a bi-unitary 
diagonalisation of the chargino mass matrix,
\begin{eqnarray*} U_-^* X U_+^{-1} = M_{ch} =  \ML m_{\chi_2^\pm} & 0 \\ 0 & m_{\chi_1^\pm}\MR \;\rm{ where} \;  U_{\pm} =
 \ML \cos \theta_{\pm} & \sin \theta_{\pm} \\ -\sin \theta_{\pm} & \cos \theta_{\pm} \MR \; \rm{if} \; \rm{det}(X) >0 
\end{eqnarray*}
 and for $\rm{det}(X) < 0\;\; U_+ \rightarrow \sigma_3 U_+$  gives us the correct matrix to diagonalise $X$ such that all masses are positive.

This leads to the chargino couplings taking the form

\BEA g_V^{ij} =  \tilde{Q}_{H_2} U_{+ \, i2} U_{+ \, j2}  - \tilde{Q}_{H_1} U_{- \, i2} U_{- \, j2}, \\
 g_A^{ij} =  \tilde{Q}_{H_2} U_{+ \, i2} U_{+ \, j2}  + \tilde{Q}_{H_1} U_{- \, i2} U_{- \, j2}.
 \EEA

For neutralino couplings the situation is similar, but in this case we have neutral 
Majorana fermions so the vector couplings vanish.   In addition,
 when we diagonalise the mass matrix numerically, we find the mixing matrix $N$ which diagonalises the 
neutralino mass matrix though $N^* M_{neut} N^{-1}$  to give diagonal masses, $m(i)$, which can be 
negative or positive.  To obtain positive masses we can then also perform a phase rotation 
$\Phi^* (N^* M_{neut} N^{-1})\Phi^{-1}$ where $(\Phi)_{jk} = (i)^{\theta(j)}\delta_{jk}$, 
where $\theta(j) = 0 (1)$ if $m(j)$ is positive (negative).  Neutralino couplings
 take then the form  
\BE g_A^{ij} =  \displaystyle\sum_{k}2 Q_k N_{ik}N_{jk}^*\;\;\mbox{where} \;\; Q_k = (0, 0, 
\tilde{Q}_{H_1}, \tilde{Q}_{H_2}, \tilde{Q}_S, 0).
   \EE

The form of the couplings $g_V^f$ and $g_A^f$ were given in \cite{King:2005jy} and are only reproduced here for 
convenience.  For the fermions of ordinary matter one has
$g^f_V = \tilde{Q}_{f_L} - \tilde{Q}_{f^c} $   and  $g^f_A = \tilde{Q}_{f_L} + \tilde{Q}_{f^c} $.  
For the exotic coloured objects we have similarly: $g^D_V = \tilde{Q}_{D} - \tilde{Q}_{\overline{D}} $   
and  $g^D_A = \tilde{Q}_{D} + \tilde{Q}_{\overline{D}} $.
For the inert Higgsinos one gets:  $g^{\tilde{H}_\alpha}_V = \tilde{Q}_{H1} - \tilde{Q}_{H2} $   and  $g^{\tilde{H}_\alpha}_A = \tilde{Q}_{H1} + \tilde{Q}_{H2}$.

The numerical values of these couplings for the benchmarks studied in this paper are shown in Tab. \ref{table:fZpCoup}.

\begin{table}[h!]
\begin{center}
\begin{tabular}{|c|c|c|c|c|}
\hline  &                \textbf{BMC Vector $g_V$} & \textbf{BMC Axial $g_A$}  & \textbf{BM4 Vector $g_V$} & \textbf{BM4 Axial $g_A$}   \\\hline
$ Z^\prime l\bar{l} $    &   $0.1108$   &  $0.4901$ &  $0.1110$   &  $0.4900$  \\
$Z^\prime \nu_l\bar{\nu_l}$ &  $0.3004$ & $0.3004$ & $0.3005$ & $0.3005$ \\
$ Z^\prime u\bar{u} $    &   $0.02630$   &  $0.3004$ &  $0.02618$   &  $0.3005$   \\
$ Z^\prime d\bar{d} $    &   $-0.1634$   &  $0.4901$ &  $-0.1633$   &  $0.4900$   \\
$Z^\prime D_i \overline{D}_i$ &  $0.1371$   &   $-0.7906$ & $0.1372$ & $-0.7906$ \\
$Z^\prime \tilde{H}_\alpha \bar{\tilde{H}}_\alpha$ &  $-0.1897$   &   $-0.7906$ & $-0.1895$   &   $-0.7906$    \\$Z^\prime \tilde{s}_\alpha \bar{\tilde{s}}_\alpha$ &  $0.7906$   &   $0.7906$  &  $0.7906$   &   $0.7906$   \\

$Z^\prime\chi^+_1\bar{\chi^+_1} $         & $-0.013565$ &  $-0.01372$   & $0.003494$     & $-0.003558$  \\
$Z^\prime\chi^+_2\bar{\chi^+_2} $         & $-0.1760$ &  $-0.7768$   & $0.1861$     & $-0.7870$  \\
$Z^\prime\chi^+_1\bar{\chi^+_2} $         & $-0.07755$ &  $-0.08396$   & $0.04450$     &$-0.03834$  \\

$Z^\prime\chi^0_1\bar{\chi^0_1} $         & $0$ &  $-0.002220$   & $0$ & $-0.0004995$      \\
$Z^\prime\chi^0_1\bar{\chi^0_2} $         & $0$ &  $-0.003660$   &  $0$     &  $0.0008260$ \\
$Z^\prime\chi^0_1\bar{\chi^0_3} $         & $0$ &  $-0.03365$   & $0$   & $-0.01560$  \\
$Z^\prime\chi^0_1\bar{\chi^0_4} $         & $0$ &  $-0.003298$   &$0$     & $0.01565$  \\
$Z^\prime\chi^0_1\bar{\chi^0_5} $         & $0$ &  $-0.03511$   & $0$    & $0$  \\
$Z^\prime\chi^0_1\bar{\chi^0_6} $         & $0$ &  $-0.001813$   &$0$     &$0.0007258$ \\
$Z^\prime\chi^0_2\bar{\chi^0_2} $         & $0$ &  $-0.006037$   & $0$     & $-0.001366$  \\
$Z^\prime\chi^0_2\bar{\chi^0_3} $         & $0$ &  $-0.05440$   &  $0$   & $0.02579$  \\
$Z^\prime\chi^0_2\bar{\chi^0_4} $         & $0$ &  $-0.005385$   & $0$     & $-0.02588$  \\
$Z^\prime\chi^0_2\bar{\chi^0_5} $         & $0$ &  $-0.0005868$   &$0$     &$0$  \\
$Z^\prime\chi^0_2\bar{\chi^0_6} $         & $0$ &  $-0.002947$   & $0$     & $-0.001222$ \\
$Z^\prime\chi^0_3\bar{\chi^0_3} $         & $0$ &  $-0.04902$   & $0$ &  $-0.4866$  \\
$Z^\prime\chi^0_3\bar{\chi^0_4} $         & $0$ &  $-0.4852$   & $0$     &  $0.4890 $ \\
$Z^\prime\chi^0_3\bar{\chi^0_5} $         & $0$ &  $-0.007955$   & $0$    & $0.03565$ \\
$Z^\prime\chi^0_3\bar{\chi^0_6} $         & $0$ &  $-0.02591$   &$0$   & $0.03806$  \\

$Z^\prime\chi^0_4\bar{\chi^0_4} $         & $0$ &  $-0.4791$   & $0$   & $-0.4901$  \\
$Z^\prime\chi^0_4\bar{\chi^0_5} $         & $0$ &  $-0.05635$   & $0$ & $0.009647$ \\
$Z^\prime\chi^0_4\bar{\chi^0_6} $         & $0$ &  $-0.05165$   & $0$     & $-0.01211$\\
$Z^\prime\chi^0_5\bar{\chi^0_5} $         & $0$ &  $-0.4569$   & $0$     &$0.4559$  \\
$Z^\prime\chi^0_5\bar{\chi^0_6} $         & $0$ &  $1.08906$   & $0$     &$1.08910 $ \\

$Z^\prime\chi^0_6\bar{\chi^0_6} $         & $0$ &  $0.52064$   &$0$ & $0.52270$  \\

\hline
\end{tabular}
\caption{Vector and axial fermion couplings to $Z^\prime_N$.}
\label{table:fZpCoup}
\end{center}
\end{table}

\end{appendix}

\end{document}